\documentclass[reprint,pre,aps,floatfix]{revtex4-2}
\usepackage{graphicx}
\usepackage{amsfonts}
\usepackage{amssymb}
\usepackage{natbib}
\usepackage{xcolor}
\usepackage{amsmath}
\usepackage{enumerate}
\usepackage{dblfloatfix}
\usepackage[percent]{overpic}
\usepackage{verbatim}
\usepackage[percent]{overpic}
\usepackage{subcaption}
\usepackage{mathtools}
\usepackage[colorlinks=true, linkcolor=blue, citecolor=blue, urlcolor=blue]{hyperref}
\usepackage{ragged2e}

\makeatletter
\long\def\@makecaption#1#2{%
  \vskip\abovecaptionskip
  \noindent\justifying
  \hangindent=0.0em\hangafter=1
  \leavevmode\normalfont#1\nobreak\hskip.5em\relax #2\par
  \vskip\belowcaptionskip}
\makeatother

\begin{document}

\title{Perturbed quantum billiards on the hyperbolic plane}

\author{Matic Orel}

\affiliation{CAMTP - Center for Applied Mathematics and Theoretical
  Physics, University of Maribor, Mladinska 3, SI-2000 Maribor, Slovenia,
  European Union}

\date{\today}

\begin{abstract}
We study the emergence of generic quantum chaos from an arithmetic billiard in
the hyperbolic plane. Starting from the triangle with an arithmetic group, we introduce
small area-preserving geometric perturbations and compute long consecutive
sequences of eigenvalues and eigenstates. The spectral statistics shows a
perturbation-dependent crossover from Poisson-like arithmetic behaviour to the
Gaussian orthogonal ensemble (GOE), with the crossover scale moving to higher
wavenumbers as the perturbation decreases. To quantify the evolution of the
nearest-neighbour spacing distribution, we compare the data with Brody and
generalized-gamma fits. The generalized-gamma form is not interpreted as a
universal crossover law. Rather, it provides a useful two-parameter diagnostic:
it allows the small- and intermediate-spacing structure to vary independently
from the large-\(s\) tail. In the small energy regime, the very-small-\(s\)
region, the bulk, and the tail are not captured by a single simple spacing
distribution. The fitting parameters therefore reveal different relaxation scales
within \(P(s)\), rather than defining a new universal law. For the weakest
perturbation we also compare the data with an effective block-GOE crossover
model. This model treats the spectrum as a superposition of weakly coupled
GOE-like blocks and provides a diagnostic of the remaining block-like structure
that persists deep into the computed spectral range.
We complement the spectral analysis with Poincar\'e--Husimi representations of
the eigenstates and quantify phase-space localization through an entropy-based
localization measure. The distribution of this measure is well described by Beta
distributions whose width decays as a power law with the wavenumber. Before the
spectral crossover, the localization statistics retains memory of the arithmetic
or near-arithmetic regime. After the GOE regime is reached, the
localization-fluctuation widths collapse onto a common decay law,
\(\sigma_{\mathrm{Beta}}\sim k^{-\eta}\), with
\(\eta\) of order \(0.37\text{--}0.38\). These results provide a combined
spectral and eigenfunction-level picture of the arithmetic-to-GOE crossover in
hyperbolic quantum billiards.
\end{abstract}

\maketitle

\section{Introduction}
\label{introduction}

One of the central questions of quantum chaos is how the statistical properties
of a quantum spectrum reflect the nature of the corresponding classical motion.
For classically integrable systems, the Berry--Tabor conjecture predicts
uncorrelated, Poissonian level statistics \cite{BerTab1977}. For generic
classically chaotic systems with time-reversal symmetry, the
Bohigas--Giannoni--Schmit conjecture \cite{BGS1984}, building on the works of Casati, Guarnieri and Valz-Gris \cite{Cas1980} predicts spectral correlations described by the Gaussian orthogonal ensemble (GOE) of random-matrix theory
\cite{Mehta,Stoe,Haake}. Billiards are among the cleanest systems in
which to test these ideas, since the classical dynamics is purely geometrical
while the quantum problem is defined by the Laplace--Beltrami operator with added boundary
conditions.
Hyperbolic billiards provide a particularly important class of examples. The
negative curvature of the hyperbolic plane makes the geodesic flow strongly
chaotic, so one would generically expect GOE spectral statistics in the quantum
problem. Arithmetic hyperbolic billiards form a striking exception to this
expectation \cite{schmit_original,aurich_level_statistics_selberg_trace}. Although their classical dynamics is chaotic, their quantum spectra can display Poisson-like rather than GOE statistics. This non-generic behaviour is tied to the arithmetic structure of the
orientation-preserving subgroup of the triangle reflection group \cite{fuchsian_group}. Through the Selberg trace formula, the spectrum of the Laplace--Beltrami operator is related to the length spectrum of periodic orbits in the billiard \cite{selberg_trace,selberg_trace_hejhal}. In arithmetic cases,
there is in addition a family of Hecke operators that commute with the
Laplace--Beltrami operator. These arithmetic symmetries impose strong
number-theoretic constraints on the quantum spectrum and are responsible for the
non-generic, Poisson-like spectral statistics
\cite{arithemtic_quantum_chaos_ref,hyperbolic_long_ref}.
The hyperbolic triangle with angles \(\pi/8\), \(\pi/2\), and \(\pi/3\),
equivalently the \((2,3,8)\) triangle group, belongs to Takeuchi's
classification of arithmetic triangle groups \cite{takeuchi_arithmetic_group_triangle}. It was
studied by Schmit as a quantum billiard and is one of the simplest billiard
examples of arithmetic quantum chaos
\cite{schmit_original,hyperbolic_triangle_schmit}.
Subsequent work by Aurich, Scheffler, and Steiner showed that the quantum
statistics of hyperbolic triangular billiards can depend sensitively on the boundary
conditions and on the associated representation of the reflection group, even
when the underlying classical length spectrum is the same
\cite{aurich_level_statistics_selberg_trace}. A natural question is therefore
how the non-generic arithmetic statistics disappear when the arithmetic
structure is broken. Earlier studies of modified Artin and Hecke billiards
observed a transition from Poissonian to GOE-like statistics and described the
crossover using Brody-type fits
\cite{Poisson_to_GOE_og,hecke_deformation}. Brody and
related phenomenological distributions have long been used to parametrize
intermediate spectral statistics \cite{Bro1973,Bro1981,Izr1990}. However, a
one-parameter interpolation forces the small-spacing level repulsion and the
large-spacing tail to evolve together. With sufficiently long spectra, these
two parts of the distribution can be resolved separately.
Progress on this question has also been limited by numerical difficulties.
Unlike in Euclidean billiards, the Laplace--Beltrami operator does not possess a simple scaling property in the spectral parameter (no direct Vergini--Saraceno method \cite{VerSar1995}), and the relevant boundary-integral kernels
involve special functions whose stable high-energy evaluation is nontrivial (see \cite{hyperbolic_comp_difficulty} for the basis functions of Artin's billiard and \cite{associated_Legendre_Q_algorithm}). On the arithmetic side, highly efficient methods based on Hecke operators are available only because of the additional number-theoretic structure \cite{fast_hecke_operators}. Long consecutive spectra for perturbed, non-arithmetic hyperbolic billiards are therefore considerably more challenging to compute.
In the present work we revisit the arithmetic-to-GOE transition for perturbations of the Schmit triangle \cite{schmit_original}. Starting from the arithmetic triangle with angles \(\pi/8\), \(\pi/2\), and \(\pi/3\), we introduce area-preserving perturbations of the angles and compute long consecutive sequences of eigenvalues. This allows us to follow the crossover over much larger spectral ranges than those used in the early numerical studies. We analyze nearest neighbour spacing distributions, sliding--window average gap ratios, spectral rigidity, and the spectral form factor. In addition to the Brody distribution, we compare the spacing distributions with
a two-parameter generalized-gamma distribution. This distribution should not be interpreted as
the true crossover distribution. Its usefulness is diagnostic: it separates the
small- and intermediate-spacing structure from the large-spacing tail, and
therefore exposes the failure of a one-parameter interpolation. The resulting
fits show that the bulk of the spacing distribution and the tail relax towards
GOE behaviour on different spectral scales.
For the weakest perturbation we additionally use an effective block-GOE
crossover model, motivated by deformed random-matrix models of symmetry
breaking \cite{Cat_map_block_GOE_perturb,Defomed_ensembles_block_GOE}, to test whether the spectrum can be viewed as several weakly coupled
subspectra whose coupling grows with increasing wavenumber. This provides a
more structured diagnostic of the slow arithmetic-to-GOE crossover than purely phenomenological spacing-distribution fits.
We also study the eigenfunctions through Poincar\'e--Husimi representations on the boundary phase space. Boundary Husimi functions provide a natural phase-space representation for billiard eigenstates
\cite{TV1995,Backer2003}, and have been used previously to study localization and scarring in hyperbolic billiards
\cite{Lepore_Heller_hyperbolic_triangle,aurich_hyperbolic_scarring}. From the Poincar\'e--Husimi density we define an entropy-based localization measure. The distributions of this measure in spectral windows are well described by Beta distributions, as in earlier studies of localization measures and chaotic
eigenstates \cite{BatRob2013A,BLR2019B}. The new point here is the
semiclassical scaling of the width of the fitted Beta distribution. Such a
power-law decay was recently identified in the \(C_3\)-symmetric Euclidean
billiard \cite{c3_arxiv_orel}. Here we show that an analogous decay appears in
perturbed hyperbolic triangles once the GOE spectral regime is reached.
The resulting picture is both spectral and eigenfunction-level. Breaking
arithmeticity does not immediately produce GOE statistics at all energies.
Instead, the crossover is perturbation-dependent and is pushed to higher
wavenumbers as the perturbation decreases. Before the crossover, both the
spacing statistics and the localization statistics retain memory of the
arithmetic or near-arithmetic regime. After the GOE regime is reached, the
spacing statistics approach the random-matrix predictions and the fluctuations
of the Poincar\'e--Husimi localization measure collapse onto a common power--law decay. Large consecutive spectra and large eigenfunction ensembles are
therefore essential for resolving the arithmetic-to-GOE crossover in hyperbolic
quantum billiards.

\section{Geometry and Classical dynamics}
\label{geometry_classical_dynamics}

\subsection{Geometry}
\label{app:geometry}

\begin{figure}
    \centering
    \includegraphics[width=\linewidth]{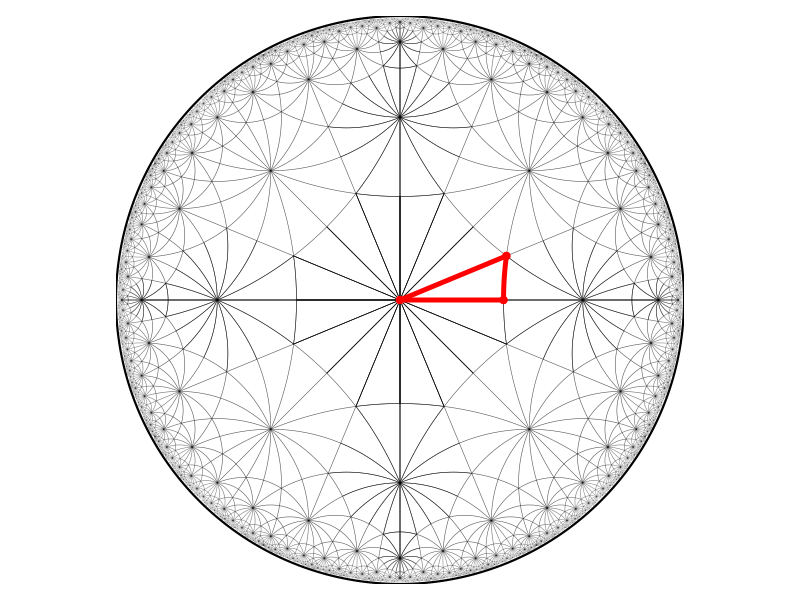}
    \caption{Finite-depth reflection tessellation of the Schmit triangle in the Poincaré disk \eqref{eq:poincare_disk}. The red triangle denotes the fundamental billiard domain, whose angles are $\pi/8$, $\pi/2$, and $\pi/3$. Since each angle is an integer submultiple of $\pi$, repeated reflections in the three geodesic sides generate a hyperbolic triangle group and tile the hyperbolic plane without gaps or overlaps. The black curves show a finite subset of this infinite tessellation; the accumulation of smaller copies near the unit circle reflects the compression of hyperbolic distance in the Poincaré disk model.}
    \label{fig:tesselation}
\end{figure}

All computations were performed in the Poincaré disk \cite{hyperbolic_triangle_schmit}

\begin{equation}
\mathbb{D}
=
\{x\in\mathbb{R}^2:\|x\|<1\},
\label{eq:poincare_disk}
\end{equation}

\noindent equipped with the hyperbolic metric $g_{ij}$ \cite{aurich_level_statistics_selberg_trace}

\begin{equation}
g_{ij}(x)
=
\frac{4}{(1-\|x\|^2)^2}\,\delta_{ij},
\qquad
ds^2
=
\frac{4\,dx\cdot dx}{(1-\|x\|^2)^2},
\label{eq:poincare_metric}
\end{equation}

\noindent where $ds$ denotes the hyperbolic line element. As a reference system we consider the arithmetic hyperbolic triangle used by Schmit (triangle $OLM$ in \cite{schmit_original}, which we will call Schmit triangle) with interior angles

\begin{equation}
(\alpha_1,\alpha_2,\alpha_3)
=
\left(
\frac{\pi}{8},
\frac{\pi}{2},
\frac{\pi}{3}
\right).
\label{eq:schmit_angles}
\end{equation}

\noindent Its area follows from the Gauss--Bonnet formula \cite{Gauss_Bonnet_triangle_area,aurich_level_statistics_selberg_trace}

\begin{equation}
\mathcal A
=
\pi-\sum_{j=1}^{3}\alpha_j
=
\frac{\pi}{24}
\approx
0.1308996939,
\label{eq:schmit_area}
\end{equation}

\noindent while the side lengths are obtained from the hyperbolic law of cosines \cite{Gauss_Bonnet_triangle_area,aurich_level_statistics_selberg_trace}

\begin{equation}
\cosh \ell_i
=
\frac{
\cos\alpha_i
+
\cos\alpha_j\cos\alpha_k
}{
\sin\alpha_j\,\sin\alpha_k
},
\qquad
(i,j,k)\ \text{cyclic},
\label{eq:hyp_cosines}
\end{equation}

\noindent giving the total arclength

\begin{equation}
L=\ell_1+\ell_2+\ell_3 \approx 1.9885116836.
\label{eq:schmit_perimeter}
\end{equation}

\noindent The billiard is seen in Fig.\ref{fig:tesselation}. To investigate the effect of geometric perturbations we consider the family of perturbations to the Schmit triangle

\begin{equation}
(\alpha_1,\alpha_2,\alpha_3)
=
\left(
\frac{\pi}{8}+\varepsilon,
\frac{\pi}{2}-\varepsilon,
\frac{\pi}{3}
\right),
\label{eq:perturbed_angles}
\end{equation}

\noindent which preserve the angle sum \eqref{eq:schmit_area} and therefore the area $\mathcal A=\frac{\pi}{24}$. The corresponding geometric parameters are listed in Table~\ref{tab:triangle_parameters}.

\begin{table}[ht]
\centering
\caption{Area-preserving perturbations of the Schmit triangle.}
\label{tab:triangle_parameters}
\begin{tabular}{ccc}
\hline
$\varepsilon$ & $\mathcal A$ & $L$ \\
\hline
$0$        & $0.1308996939$ & $1.9885116836$ \\
$10^{-5}$  & $0.1308996939$ & $1.9884970016$ \\
$10^{-4}$  & $0.1308996939$ & $1.9883648936$ \\
$10^{-3}$  & $0.1308996939$ & $1.9870467877$ \\
\hline
\end{tabular}
\end{table}

\subsection{Classical billiard map}
\label{app:classical_map}

For the classical comparison we used the exact billiard map in the Poincaré
disk model. The billiard boundary consists of three hyperbolic geodesic sides.
In the disk model, such geodesics are represented either by Euclidean diameters
of the unit disk or by Euclidean circles orthogonal to the unit circle. Thus
both the boundary arcs and the free-flight trajectories admit closed-form
Euclidean descriptions.

Let \(p\in\mathbb D\) be a point on a trajectory and let \(v\) be its Euclidean
tangent direction. If the corresponding hyperbolic geodesic is not a diameter,
it is represented by a Euclidean circle orthogonal to \(\partial\mathbb D\).
Its center \(c\) is determined by the two linear conditions \cite{fuchsian_group}

\begin{equation}
    c\cdot p=\frac{1+|p|^2}{2},
    \qquad
    c\cdot v=p\cdot v .
    \label{eq:geodesic_circle_center}
\end{equation}

\noindent The first condition imposes orthogonality of the geodesic circle to the unit
circle, while the second imposes that the circle is tangent to the prescribed
direction \(v\) at \(p\). The Euclidean radius of the geodesic circle is then \cite{fuchsian_group}

\begin{equation}
    R_g=\sqrt{|c|^2-1}.
    \label{eq:geodesic_circle_radius}
\end{equation}

The diameter case is treated separately as a Euclidean line through the origin.

\begin{figure}
    \centering
    \includegraphics[width=\linewidth]{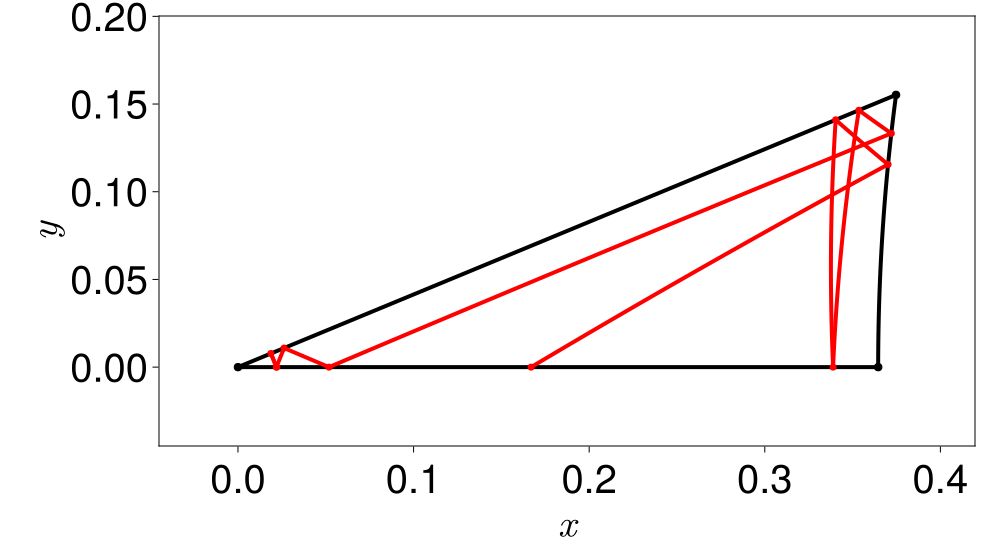}
    \caption{Example classical trajectory in Schmit's hyperbolic triangle. The
    black curve shows the billiard boundary in the Poincaré disk model, and the
    red arcs show several specular reflections of a geodesic trajectory. The two
    sides emanating from the origin are radial geodesics, while the third side
    is a circular arc orthogonal to the unit disk boundary.}
    \label{fig:classical_bounces}
\end{figure}

The next collision is obtained by intersecting the current geodesic with all
three boundary sides and selecting the admissible intersection lying first in
the forward direction of motion. Since all curves involved are Euclidean lines
or circles, this requires only explicit line--line, line--circle, and
circle--circle intersections. The previous collision point is excluded by a
small tolerance. Thus the classical map is evaluated without numerical root finding for the
collision points. An example of such a reflected geodesic trajectory is shown in
Fig.~\ref{fig:classical_bounces}.

At the collision point \(q\), the incoming Euclidean tangent vector
\(v_{\mathrm{in}}\) is reflected specularly with respect to the Euclidean
outward unit normal \(n(q)\) of the boundary,

\begin{equation}
    v_{\mathrm{out}} = v_{\mathrm{in}} - 2\bigl(v_{\mathrm{in}}\cdot n(q)\bigr)n(q).
    \label{eq:classical_reflection}
\end{equation}

\noindent This is equivalent to hyperbolic specular reflection because the Poincaré disk
metric is conformal to the Euclidean metric and therefore preserves angles.

As a numerical check of the classical instability, we computed the maximal
Lyapunov exponent using the standard Benettin algorithm
\cite{benettin_normalization_1,benettin_normalization_2}. The calculation was performed on a grid of initial
conditions in the phase space. For all perturbations considered here, the billiard remains a geodesic-sided hyperbolic
triangle. Numerically, we find \(\lambda_{\max}\simeq1\) throughout the entire phase space for all
values of \(\epsilon\). Thus the perturbation breaks the arithmetic
reflection-group structure, but does not change the local hyperbolic instability
scale.

\section{Quantum problem}
\label{quantum_problem}

We study the Dirichlet spectrum of the Laplace--Beltrami operator on the
hyperbolic billiard domain \(\Omega\subset\mathbb D\). With the sign convention
used throughout this work, the eigenvalue problem is

\begin{equation}
-\Delta_H \psi_n
=
\lambda_n \psi_n,
\quad
\psi_n|_{\partial\Omega}=0,
\quad
\lambda_n=k_n^2+\frac14.
\label{eq:dirichlet_problem}
\end{equation}

In the Poincar\'e disk metric \eqref{eq:poincare_metric}, the Laplace--Beltrami
operator is \cite{schmit_original}

\begin{equation}
\Delta_H
=
\frac{(1-r^2)^2}{4}
\left(
\frac{\partial^2}{\partial r^2}
+
\frac{1}{r}\frac{\partial}{\partial r}
+
\frac{1}{r^2}\frac{\partial^2}{\partial\theta^2}
\right),
\label{eq:laplace_beltrami_polar}
\end{equation}

\noindent or equivalently,

\begin{equation}
\Delta_H
=
\frac{(1-|x|^2)^2}{4}
\Delta_E,
\label{eq:laplace_beltrami_cartesian}
\end{equation}

\noindent where \(\Delta_E\) is the Euclidean Laplacian. Thus the semiclassical parameter is
\(k\), and the high-energy limit corresponds to \(k\to\infty\). For Dirichlet eigenvalues of the Laplace--Beltrami operator, written as $\lambda=k^2+\frac14$, the counting function satisfies \cite{Weyl1911,aurich_level_statistics_selberg_trace,hyperbolic_long_ref}

\begin{equation}
N(k) = \frac{\mathcal A}{4\pi} \left(k^2+\frac14\right) - \frac{\mathcal L}{4\pi} \sqrt{k^2+\frac14} + C_{\rm cc}.
\label{eq:weyl_k}
\end{equation}

\noindent where $C_{cc}$ is the corner and curvature correction and $\mathcal A,\mathcal L$ are the hyperbolic area and arclength \eqref{eq:schmit_area}, respectively. Since the area is fixed throughout the family \eqref{eq:perturbed_angles}, the leading Weyl coefficient is unchanged and only the perimeter correction and $C_{cc}$ vary with $\varepsilon$.

The Green function used in the boundary-integral formulation is the resolvent
kernel for the equation \cite{aurich_equation_wavefunction_hyperbolic}

\begin{equation}
\left(
\Delta_H+\frac14+k^2
\right)
G_k(x,y)
=
\delta(x-y),
\label{eq:green_equation}
\end{equation}

\noindent and can be written in terms of the Legendre function of the second kind as \cite{aurich_equation_wavefunction_hyperbolic}

\begin{equation}
G_k(x,y)
=
\frac{1}{2\pi}
Q_{\nu}\!\left(\cosh\chi(x,y)\right),
\qquad
\nu=-\frac12+ik.
\label{eq:hyperbolic_green_main}
\end{equation}

\noindent Here \(\chi(x,y)=d_H(x,y)\) is the hyperbolic distance. In the Poincar\'e disk it
is given explicitly by \cite{aurich_equation_wavefunction_hyperbolic}

\begin{equation}
\cosh\chi(x,y)
=
1+
\frac{2|x-y|^2}
{(1-|x|^2)(1-|y|^2)}.
\label{eq:hyperbolic_distance_main}
\end{equation}

\noindent This closed form is one of the main numerical advantages of the disk model: all
matrix entries of the boundary-integral formulation can be expressed in terms of
the Euclidean coordinates of the boundary nodes. Eq.\eqref{eq:hyperbolic_green_main} has the same logarithmic singular structure as the Euclidean Helmholtz double-layer kernel \cite{backer_BIM}. Indeed, as \(\chi\to0\) (see \cite{NIST:DLMF}),

\begin{equation}
Q_\nu(\cosh\chi) = -P_\nu(\cosh\chi)\log\frac{\chi}{2} + B_\nu(\chi),
\label{eq:Q_log_structure_main}
\end{equation}

\noindent where \(B_\nu\) is smooth. Hence the normal derivative of the Green function can
be split into a logarithmic part and a smooth remainder. This is treated by a corner graded
Kress analytic split in a periodic boundary parameter \cite{kress}. The details of the
split, including the conformal correction to the Euclidean diagonal term, are given in
Appendix~\ref{app:hyp_kress_split}.

We now reformulate the Dirichlet eigenvalue problem as a boundary-integral
equation \cite{kress_book}. For a boundary density \(\mu\), define the
double-layer potential

\begin{equation}
\mathcal D_k\mu(x)
=
\int_{\partial\Omega}
\partial_{n_y^H}G_k(x,y)\,
\mu(y)\,ds_H(y),
\qquad x\in\Omega .
\label{eq:double_layer_potential_main}
\end{equation}

\noindent This expression has the advantage that it satisfies the differential
equation in the interior automatically. Indeed, for \(x\in\Omega\), the source
point \(y\) lies on \(\partial\Omega\), and therefore the singularity of the
Green function is never reached in the \(x\)-variable. Hence

\begin{equation}
\left(
\Delta_H+\frac14+k^2
\right)
\mathcal D_k\mu(x)
=
0,
\qquad x\in\Omega .
\label{eq:dlp_solves_pde}
\end{equation}

\noindent Thus the two-dimensional Dirichlet eigenvalue problem is reduced to imposing the
interior boundary limit giving the same jump relation as the Euclidean metric \cite{kress_book,aurich_equation_wavefunction_hyperbolic}

\begin{equation}
\lim_{\substack{x\to x_0\\ x\in\Omega}}
\mathcal D_k\mu(x) = -\frac12\mu(x_0) + K_k\mu(x_0), \quad x_0\in\partial\Omega,
\label{eq:dlp_jump_relation}
\end{equation}

where

\begin{equation}
K_k\mu(x) = \operatorname{p.v.}
\int_{\partial\Omega}
\partial_{n_y^H}G_k(x,y)\,
\mu(y)\,ds_H(y).
\label{eq:boundary_double_layer_operator}
\end{equation}

\noindent Here \(K_k\) denotes the principal-value boundary operator obtained from the boundary limit of the double-layer potential \eqref{eq:double_layer_potential_main}. The Dirichlet boundary condition \(\psi|_{\partial\Omega}=0\) therefore becomes

\begin{equation}
\left(
-\frac12 I+K_k
\right)\mu=0.
\label{eq:bie_unscaled}
\end{equation}

\noindent Equivalently, after multiplying by \(-2\), we solve

\begin{equation}
A(k)\mu=0,
\qquad
A(k)=I-D(k),
\qquad
D(k)=2K_k .
\label{eq:fredholm_main}
\end{equation}

\noindent Although the operator is hyperbolic, the Poincar\'e disk metric is conformal to
the Euclidean metric. Writing

\begin{equation}
g_H=\lambda^2 g_E,
\qquad
\lambda(y)=\frac{2}{1-|y|^2},
\end{equation}

\noindent we have

\begin{equation}
ds_H=\lambda\,ds_E,
\qquad
n^H=\lambda^{-1}n^E,
\label{eq:conformal_scaling_main}
\end{equation}

\noindent where $n^H$ and $n^E$ are the outward hyperbolic and Euclidean unit normal at a point on the $\partial \Omega$, respectively. Hence the normal derivative and arclength factors cancel,

\begin{equation}
\partial_{n_y^H}G_k(x,y)\,ds_H(y)
=
\partial_{n_y^E}G_k(x,y)\,ds_E(y).
\label{eq:normal_arclength_cancellation_main}
\end{equation}

\noindent Consequently, the boundary operator can be assembled in the equivalent form

\begin{equation}
K_k\mu(x)
=
\operatorname{p.v.}
\int_{\partial\Omega}
\partial_{n_y^E}G_k(x,y)\,
\mu(y)\,ds_E(y),
\label{eq:boundary_double_layer_euclidean_form}
\end{equation}

\noindent where all hyperbolic geometry enters through the Green function
\(G_k(x,y)=G_k(\chi(x,y))\). This is the form used in the numerical
discretization. The N\"ystrom discretized boundary equation \eqref{eq:fredholm_main} has the
form

\begin{equation}
A_N(k)\mu_N=0,
\label{eq:discrete_fredholm_main}
\end{equation}

\noindent where \(A_N(k)\) is a dense \(N\times N\) matrix depending nonlinearly on the
spectral parameter \(k\). The eigenvalues are then found with Beyn's contour-integral method applied to the matrix problem \eqref{eq:discrete_fredholm_main} (for details see Appendix~\ref{app:beyn_method}).

\section{Spectral Statistics}
\label{spectral_statistics}

\subsection{Nearest-neighbour level spacings (NNLS)}
\label{subsec:nnls}

We first study the nearest-neighbour level-spacing distribution. The spectra are
unfolded using the Weyl law \eqref{eq:weyl_k}. For consecutive eigenvalues
\(k_n<k_{n+1}\), we define the unfolded spacings by

\begin{equation}
s_n
=
N(k_{n+1})-N(k_n),
\label{eq:unfolded_spacings}
\end{equation}

\noindent so that \(\langle s\rangle=1\). As reference distributions we use the Poisson
distribution,

\begin{equation}
P_{\rm P}(s)=e^{-s},
\label{eq:poisson_nnls}
\end{equation}

\noindent and the GOE Wigner surmise,

\begin{equation}
P_{\rm GOE}(s)
=
\frac{\pi}{2}s
\exp\left(-\frac{\pi s^2}{4}\right).
\label{eq:goe_nnls}
\end{equation}

The Poisson distribution has no level repulsion and an exponential tail, whereas
the GOE distribution has linear level repulsion at the origin and a Gaussian
tail. To quantify the crossover between these two limits, we compare the data
with the Brody distribution

\begin{equation}
P_{\rm B}(s;\beta)
=
(\beta+1)b\,s^\beta
\exp\left(-b s^{\beta+1}\right),
\
b = \left[ \Gamma\left( \frac{\beta+2}{\beta+1} \right) \right]^{\beta+1}.
\label{eq:brody_nnls}
\end{equation}

\begin{figure}
    \centering
    \includegraphics[width=\linewidth]{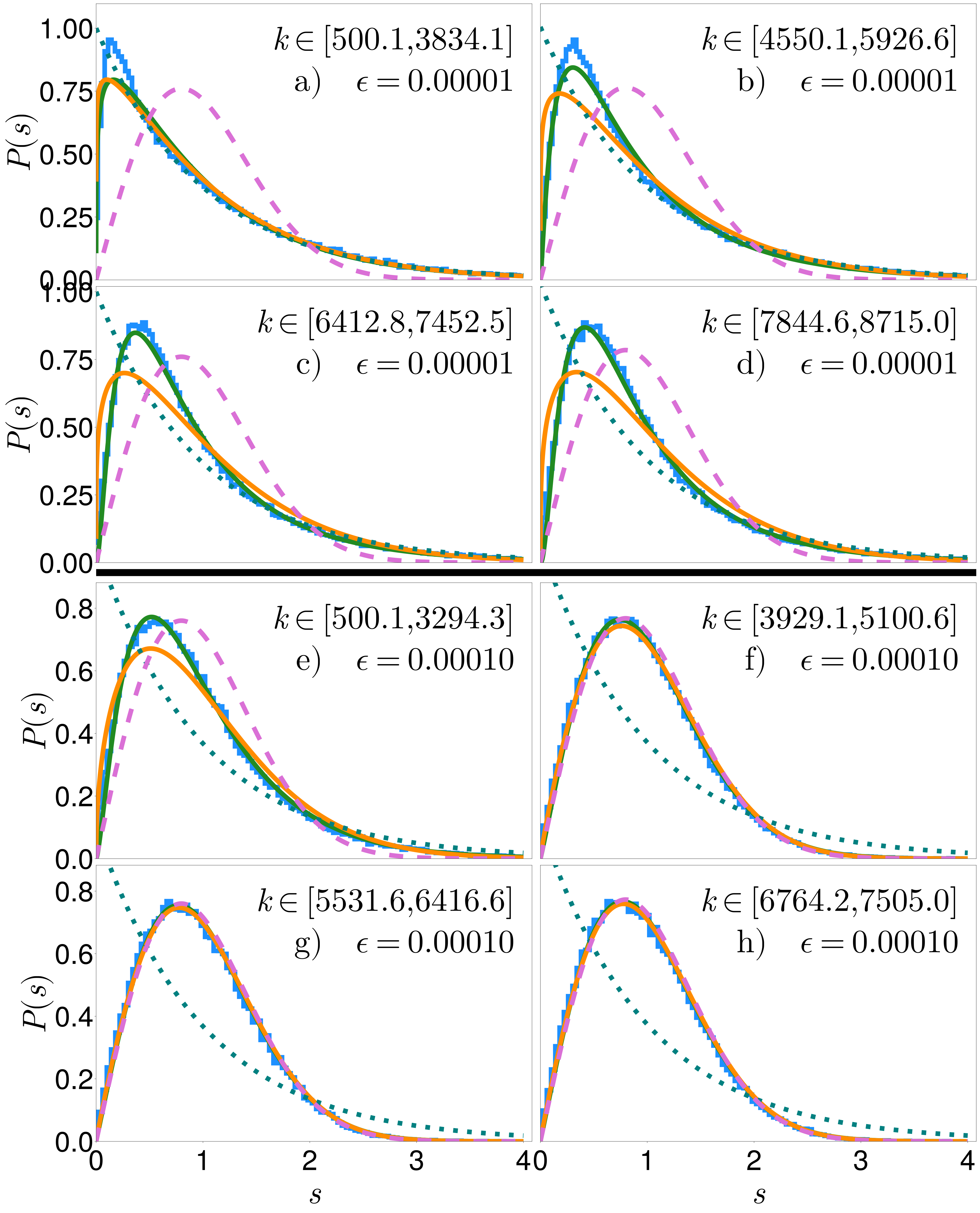}
    \caption{
    Nearest-neighbour spacing distributions for the weak perturbations
    \(\epsilon=10^{-5}\) and \(\epsilon=10^{-4}\), shown in four spectral
    windows for each perturbation. Panels (a)--(d) correspond to
    \(\epsilon=10^{-5}\) with
    \(k\in[500.1,3834.1]\), \(k\in[4550.1,5926.6]\),
    \(k\in[6412.8,7452.5]\), and \(k\in[7844.6,8715.0]\),
    respectively. Each of these panels contains \(1.5\times10^5\) consecutive
    eigenvalues, corresponding to \(149999\) unfolded spacings. Panels
    (e)--(h) correspond to \(\epsilon=10^{-4}\) with
    \(k\in[500.1,3294.3]\), \(k\in[3929.1,5100.6]\),
    \(k\in[5531.6,6416.6]\), and \(k\in[6764.2,7505.0]\),
    respectively. Each of these panels contains \(1.1\times10^5\) consecutive
    eigenvalues, corresponding to \(109999\) unfolded spacings. The blue
    histograms show the unfolded numerical spacings. The dotted curve is the
    Poisson prediction, the dashed curve is the GOE Wigner surmise, and the
    solid curves show Brody (orange) and generalized-gamma (green) fits. For
    \(\epsilon=10^{-5}\), the spectrum remains close to Poisson over a large
    spectral range, and the lowest window is not fully captured by either fit.
    At larger \(k\), the generalized-gamma distribution gives a better global
    description than the Brody distribution. For \(\epsilon=10^{-4}\), the
    spacing distribution flows steadily towards GOE.
    }
    \label{fig:P_s_1e-5_1e-4}
\end{figure}

\begin{figure}
    \centering
    \includegraphics[width=\linewidth]{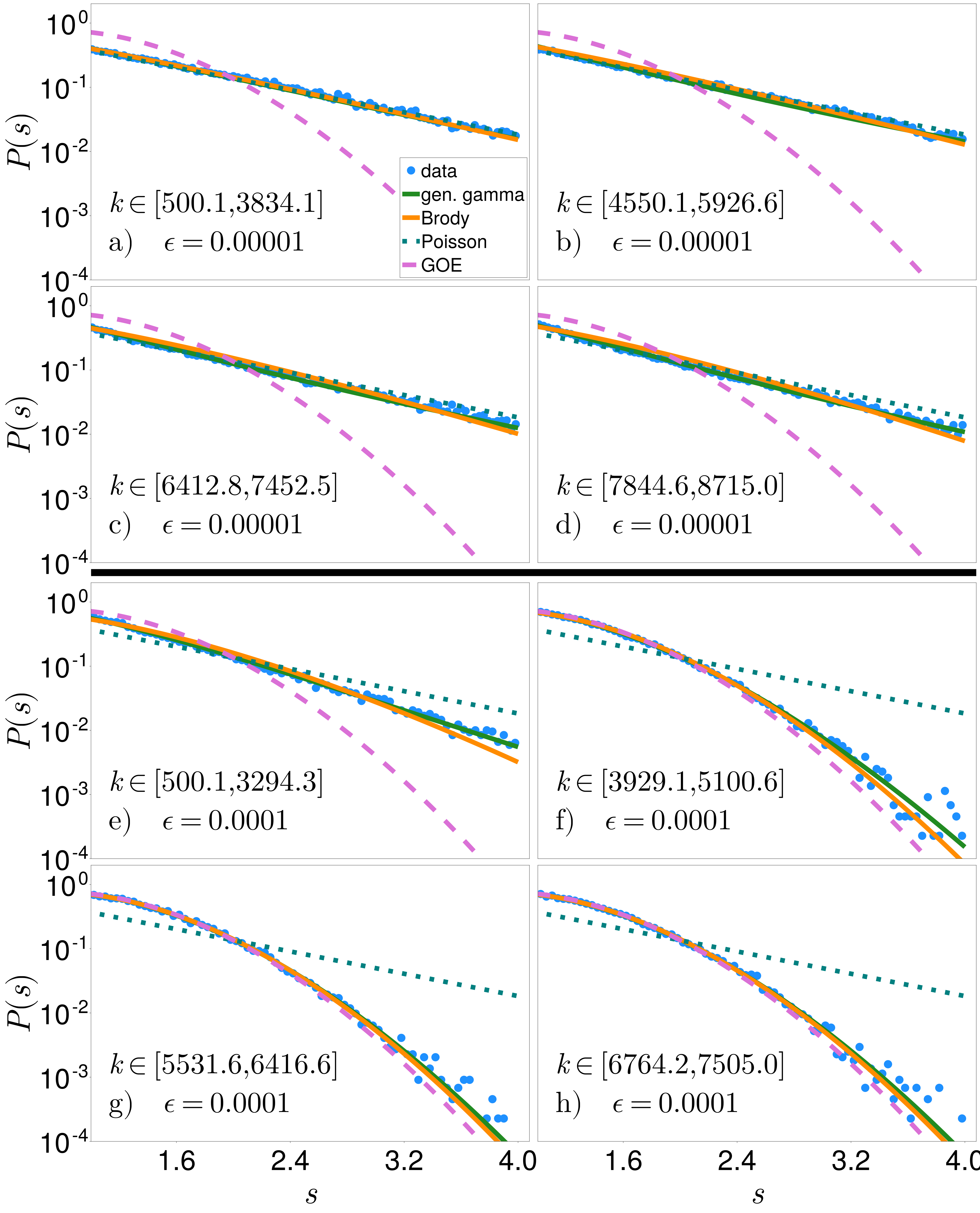}
    \caption{
    Same nearest-neighbour spacing distributions as in
    Fig.~\ref{fig:P_s_1e-5_1e-4}, shown on a logarithmic vertical scale to
    emphasize the large-spacing tail. Panels (a)--(d) correspond to
    \(\epsilon=10^{-5}\) with
    \(k\in[500.1,3834.1]\), \(k\in[4550.1,5926.6]\),
    \(k\in[6412.8,7452.5]\), and \(k\in[7844.6,8715.0]\),
    respectively, each using \(1.5\times10^5\) consecutive eigenvalues
    (\(149999\) unfolded spacings). Panels (e)--(h) correspond to
    \(\epsilon=10^{-4}\) with
    \(k\in[500.1,3294.3]\), \(k\in[3929.1,5100.6]\),
    \(k\in[5531.6,6416.6]\), and \(k\in[6764.2,7505.0]\),
    respectively, each using \(1.1\times10^5\) consecutive eigenvalues
    (\(109999\) unfolded spacings). For \(\epsilon=10^{-5}\), the tail remains
    close to exponential over much of the accessible range. For
    \(\epsilon=10^{-4}\), the tail progressively crosses over towards the
    faster GOE-like decay. The generalized-gamma fit (green) captures this tail
    evolution more accurately than the one-parameter Brody fit (orange). The dotted blue line is Poisson
    and the dashed purple line is GOE.
    }
    \label{fig:P_s_tails_1e-5_1e-4}
\end{figure}

\noindent Here \(\beta=0\) gives the Poisson distribution, while \(\beta=1\) gives the
GOE Wigner surmise. However, in the present data the small-\(s\) behaviour and
the large-\(s\) tail do not always evolve at the same rate (see Figs.\ref{fig:P_s_1e-5_1e-4},\ref{fig:P_s_tails_1e-5_1e-4}). We therefore also compare the data with a two-parameter generalized-gamma distribution \cite{stacy_generalized_gamma},

\begin{equation}
P_{\rm GG}(s;\beta,\gamma)
=
C\,s^\beta
\exp\left(-a s^\gamma\right),
\label{eq:generalized_gamma_nnls}
\end{equation}

\noindent where \(a\) and \(C\) are fixed by normalization and by the condition
\(\langle s\rangle=1\),

\begin{equation}
a=\left[\frac{\Gamma\left((\beta+2)/\gamma\right)}{\Gamma\left((\beta+1)/\gamma\right)}\right]^\gamma,
\quad
C=\frac{ \gamma a^{(\beta+1)/\gamma} }{ \Gamma\left((\beta+1)/\gamma\right) }.
\label{eq:generalized_gamma_norm_nnls}
\end{equation}

In this parametrization, the Poisson distribution corresponds to
\((\beta,\gamma)=(0,1)\), while the GOE Wigner surmise corresponds to
\((\beta,\gamma)=(1,2)\). For an exact generalized-gamma distribution, \(\beta\) controls the power-law
behaviour near the origin, while \(\gamma\) controls the decay of the tail. In
the present spectra, however, the data are not exactly of generalized-gamma
form. We therefore interpret the fitted parameters only as effective diagnostics:
\(\beta_{\mathrm{GG}}\) mainly tracks the small- and intermediate-spacing
structure of the fitted curve, while \(\gamma_{\mathrm{GG}}\) tracks the
large-spacing tail. For the case of $\epsilon=0.0$ and $1e-3$ we calculated about $210000$ eigenvalues, for the case of $5e-5$ we used $372000$ eigenvalues, for $\epsilon=1e-4$ we used $583000$ eigenvalues and finally for $\epsilon=1e-5$ we used $782000$ eigenvalues.

\begin{figure}
    \centering
    \includegraphics[width=0.9\linewidth]{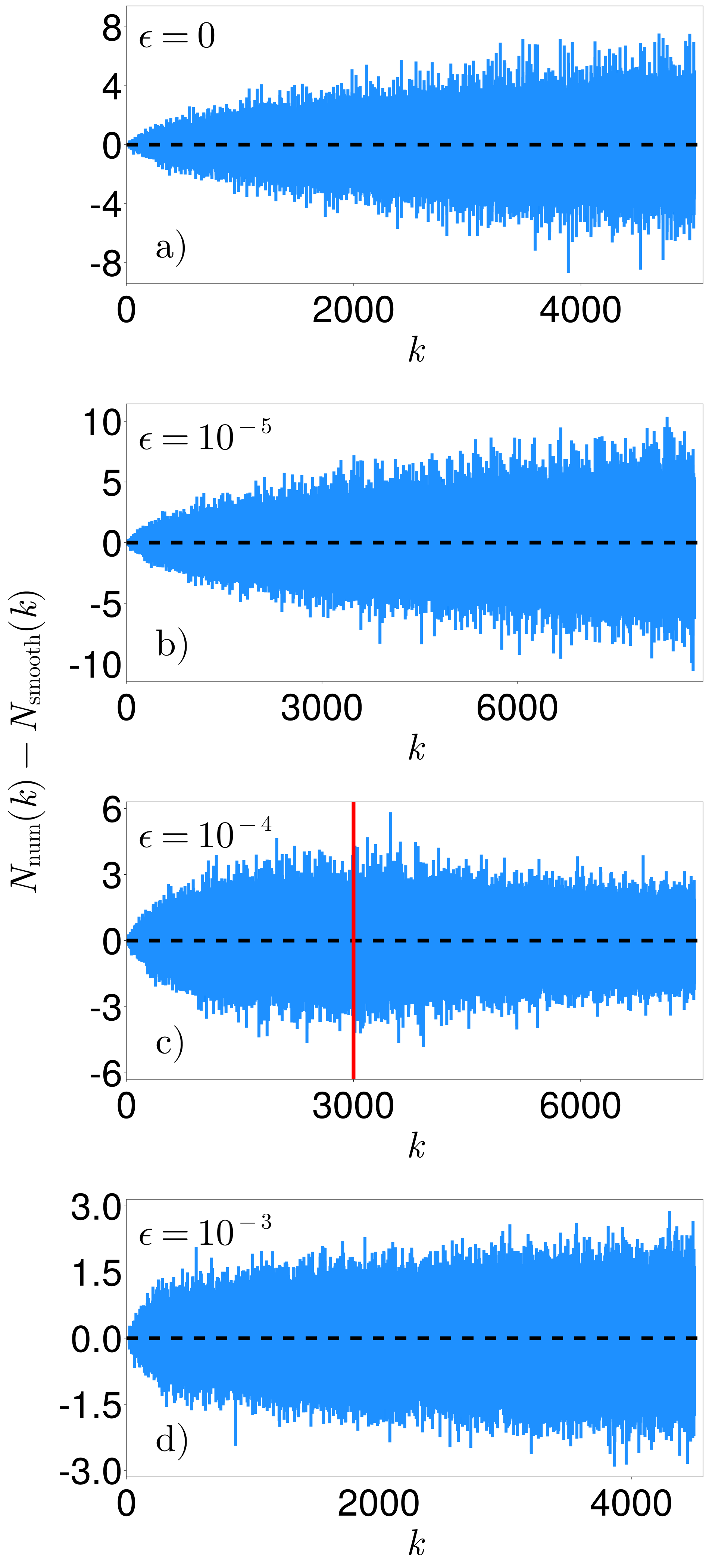}
    \caption{
    Deviation of the numerical counting function from the Weyl law
    \eqref{eq:weyl_k}. The fluctuations remain centered around zero and show no
    systematic drift, indicating that the computed spectra are complete on the
    plotted intervals. For \(\epsilon=10^{-4}\), the fluctuations decrease after
    approximately \(k\simeq3000\) (red vertical line in panel c)), consistent with the simultaneous approach of
    the local spacing statistics and the sliding--window average gap ratio towards GOE as seen in Fig.\ref{fig:average_r}.
    }
    \label{fig:weyl}
\end{figure}

\begin{figure}
    \centering
    \includegraphics[width=\linewidth]{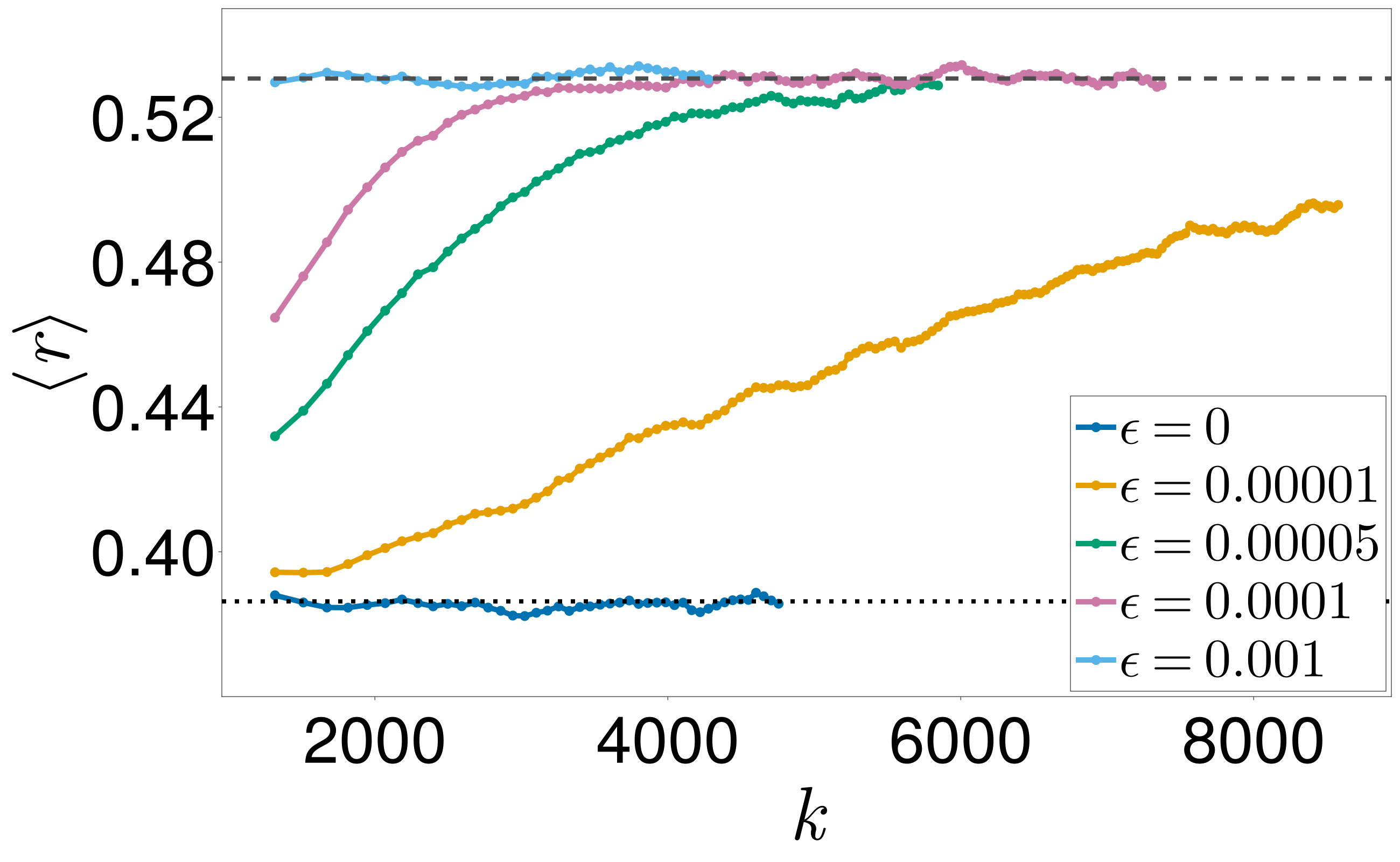}
    \caption{
    Sliding--window average gap ratio \(\langle r\rangle\) for the Schmit triangle and its
    perturbations. For each value of \(\epsilon\), the average is computed in
    sliding windows of \(4\times10^4\) consecutive eigenvalues, advanced by
    \(5\times10^3\) eigenvalues between neighbouring points. Each point therefore
    contains \(39998\) local gap ratios. The horizontal dotted line marks the
    Poisson value \(\langle r\rangle_{\mathrm{P}}=2\log2-1\approx0.38629\), while
    the dashed line marks the infinite-dimensional GOE value
    \(\langle r\rangle_{\mathrm{GOE}}\approx0.5307\) \cite{AtaBogGirRou2013}. The unperturbed Schmit
    triangle remains close to the Poisson value, whereas the \(\epsilon=10^{-3}\)
    perturbation rapidly approaches GOE statistics. The intermediate perturbations
    show a scale-dependent crossover; in particular, for \(\epsilon=10^{-4}\),
    \(\langle r\rangle\) approaches the GOE value after approximately
    \(k\simeq3000\), consistent with the decrease of the Weyl-law fluctuations and
    the evolution of the spacing distributions.
    }
    \label{fig:average_r}
\end{figure}

Before analyzing the spacings, we checked the completeness of the spectra by
comparing the numerical counting function with Weyl's law. The deviations are
shown in Fig.~\ref{fig:weyl}. In all cases the fluctuations remain centered
around zero, with no systematic drift. This indicates that no levels are missed. 
This check is particularly important in the present problem,
because the conclusions about the crossover from arithmetic to generic
statistics rely on long consecutive spectral sequences (see Figs.\ref{fig:average_r},\ref{fig:main_power_law}).

The spacing distributions for the weak perturbations
\(\epsilon=10^{-5}\) and \(\epsilon=10^{-4}\) are shown in
Fig.~\ref{fig:P_s_1e-5_1e-4}. For \(\epsilon=10^{-5}\), the distribution remains
close to Poisson over a very large part of the computed spectrum. In the smallest $k$
window, the distribution cannot be fully described by either the Brody or the
generalized-gamma fit: the small-spacing region and the tail are both too
structured to be represented by a simple interpolation. At larger \(k\),
however, the generalized-gamma curve provides a more flexible diagnostic description of the changing shape, although it should not be interpreted as the underlying spacing law. This shows that, when the spectrum is still far from GOE, neither Brody nor the generalized-gamma form provides a genuine crossover law. This result improves upon the pioneering results of \cite{Poisson_to_GOE_og,hecke_deformation} where they found a Brody distribution fit for a slightly perturbed Artin's billiard, where they only used $2000$ eigenvalues. For \(\epsilon=10^{-4}\), the crossover is much faster. The peak of \(P(s)\) moves faster away from the origin, the small spacings are progressively suppressed, and
the distribution steadily approaches the GOE curve.

The tail behaviour is shown in the logarithmic version of the same data in
Fig.~\ref{fig:P_s_tails_1e-5_1e-4}. For \(\epsilon=10^{-5}\), the tail remains
close to exponential for a long range of \(k\), even when the bulk has already
started to deviate from the Poisson form. For \(\epsilon=10^{-4}\), the tail
crosses over towards the faster GOE-like decay. This is the clearest indication
that the Brody distribution is too restrictive: forcing the level-repulsion
exponent and the tail exponent to evolve together gives a poor global
description. The generalized-gamma fit is systematically better because it
separates these two effects.

\begin{figure}
    \centering
    \includegraphics[width=\linewidth]{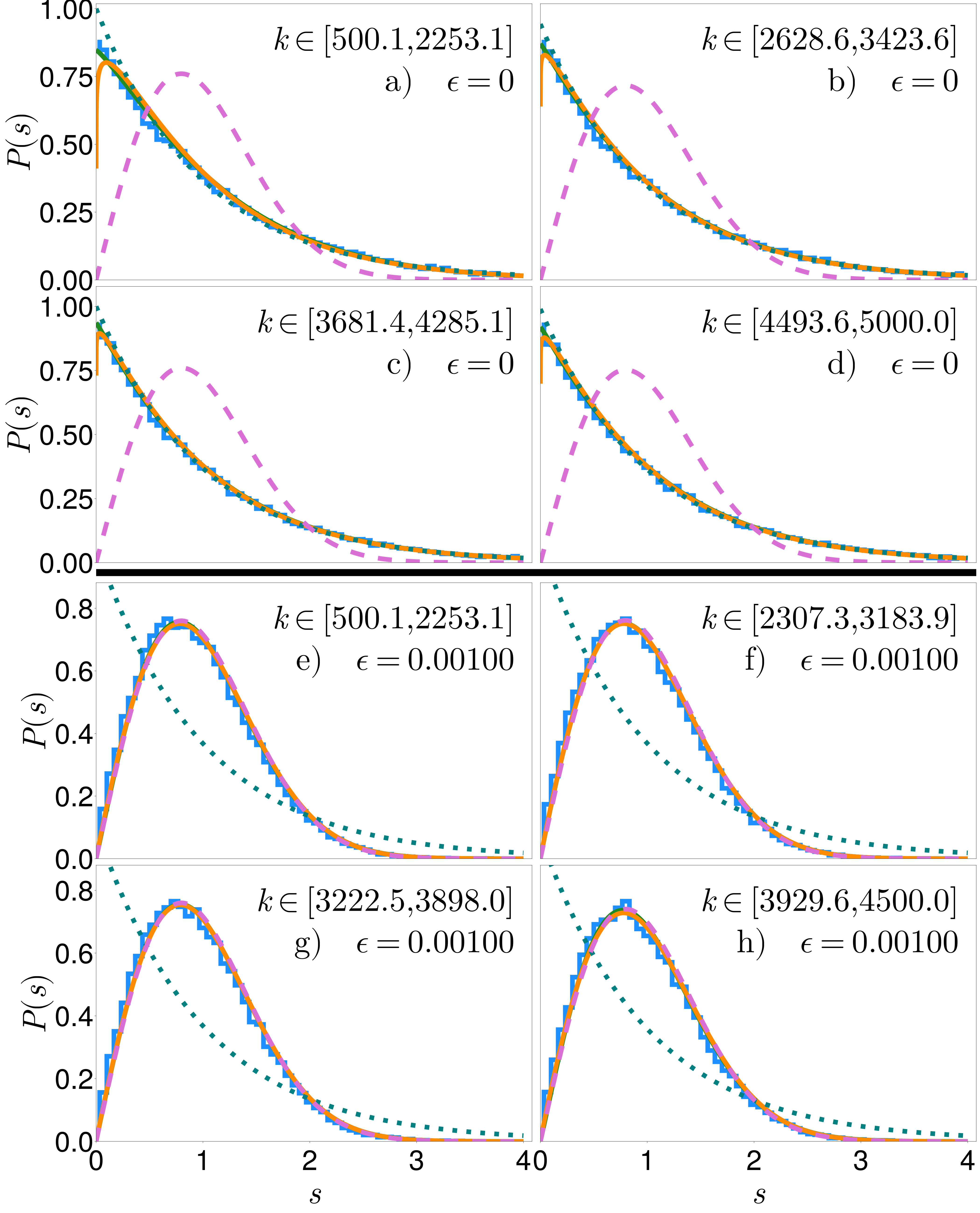}
    \caption{
    Nearest-neighbour spacing distributions for the unperturbed Schmit triangle
    \(\epsilon=0\) and the perturbed triangle \(\epsilon=10^{-3}\), shown in
    four spectral windows for each case. Panels (a)--(d) correspond to
    \(\epsilon=0\) with
    \(k\in[500.1,2253.1]\), \(k\in[2628.6,3423.6]\),
    \(k\in[3681.4,4285.1]\), and \(k\in[4493.6,5000.0]\),
    respectively. Panels (e)--(h) correspond to \(\epsilon=10^{-3}\) with
    \(k\in[500.1,2253.1]\), \(k\in[2307.3,3183.9]\),
    \(k\in[3222.5,3898.0]\), and \(k\in[3929.6,4500.0]\),
    respectively. Each panel contains \(5\times10^4\) consecutive eigenvalues,
    corresponding to \(49999\) unfolded spacings. The arithmetic case remains
    essentially Poissonian (dotted blue line) throughout the computed spectral range. In contrast,
    for \(\epsilon=10^{-3}\), the distribution is already close to GOE (dashed purple line) in the
    first spectral window and remains close to GOE as \(k\) is increased. The orange solid line is the best fitting Brody and the green solid line the best fitting generalized-gamma.
    }
    \label{fig:P_s_1e-3_0.0}
\end{figure}

Fig.~\ref{fig:P_s_1e-3_0.0} compares the unperturbed Schmit triangle
\(\epsilon=0\) with the larger perturbation \(\epsilon=10^{-3}\). The
unperturbed arithmetic case remains essentially Poissonian throughout the
computed range, in agreement with Schmit's original observation for the
tessellating triangle \cite{schmit_original}. In contrast, for \(\epsilon=10^{-3}\), the spacing
distribution is already very close to GOE in the first window and remains stable
as \(k\) is increased. Thus a perturbation of size \(10^{-3}\) is already
sufficient to destroy the arithmetic group spacing statistics on the spectral scales
studied here.

\begin{figure}
    \centering
    \includegraphics[width=\linewidth]{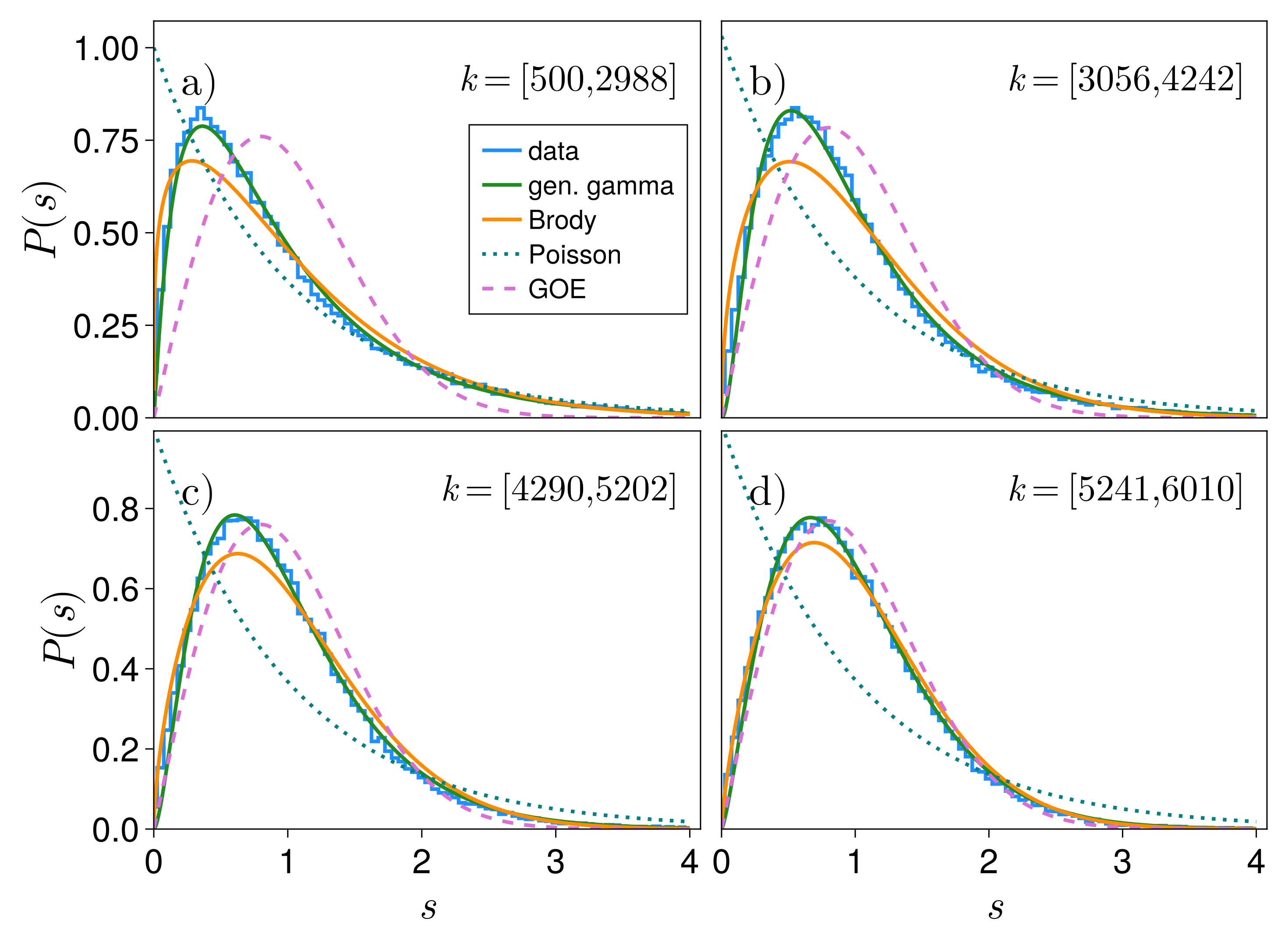}
    \includegraphics[width=\linewidth]{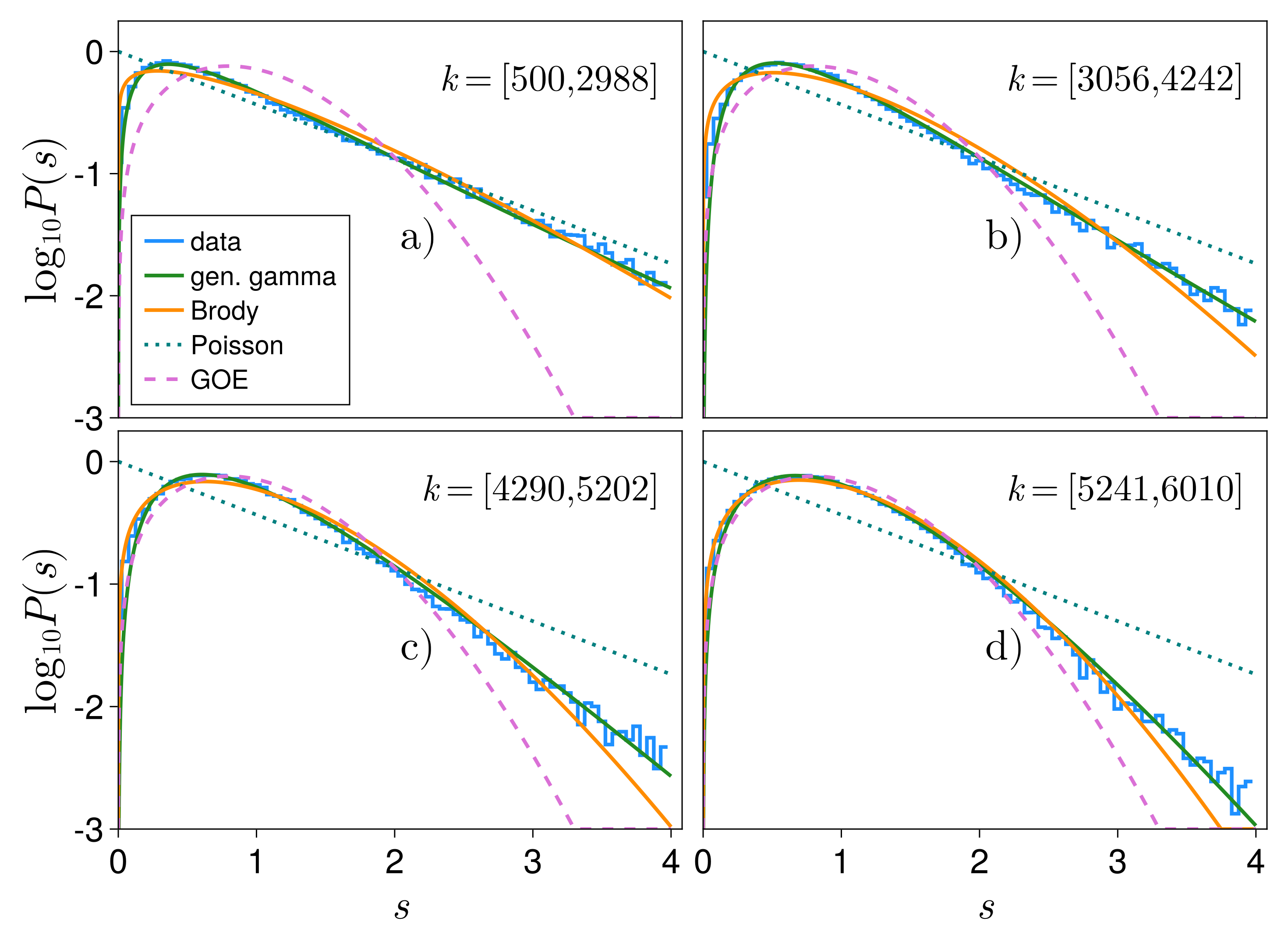}
    \caption{
    NNLS for the perturbed hyperbolic triangle with $\epsilon=5\times10^{-5}$. Each spectral window contains $90000$ eigenvalues. The upper four panels show $P(s)$ for increasing windows in $k$, while the lower four panels show the same data on a logarithmic scale, emphasizing the large-$s$ tail. The numerical histogram is compared with Poisson and GOE predictions, together with Brody and generalized-gamma fits. The generalized-gamma distribution gives a substantially better (but not perfect) description than the Brody distribution, since it can adjust the level repulsion and tail decay independently. It captures the bulk of the distribution and the tail qualitatively well, but does not reproduce the very-small-$s$ behaviour accurately. The comparison shows that the spacing distribution approaches the GOE limit through a gradual deformation of both the small-spacing region and the tail.
    }
    \label{fig:P_s_5e-5}
\end{figure}

The same transition is visible in the sliding--window average gap ratio
\(\langle r\rangle\), shown in Fig.~\ref{fig:average_r}. For the unperturbed
case, \(\langle r\rangle\) stays close to the Poisson value. For
\(\epsilon=10^{-3}\), it rapidly approaches the GOE value. The intermediate
perturbations show a scale-dependent crossover. In particular, for
\(\epsilon=10^{-4}\), the Weyl fluctuations decrease after approximately
\(k\simeq3000\), and over the same range the sliding--window average gap ratio
\(\langle r\rangle\) moves towards the GOE value. For a more representative picture
see the perturbation \(\epsilon=5\times10^{-5}\) in Fig.~\ref{fig:P_s_5e-5}.

\begin{figure}
    \centering
    \includegraphics[width=\linewidth]{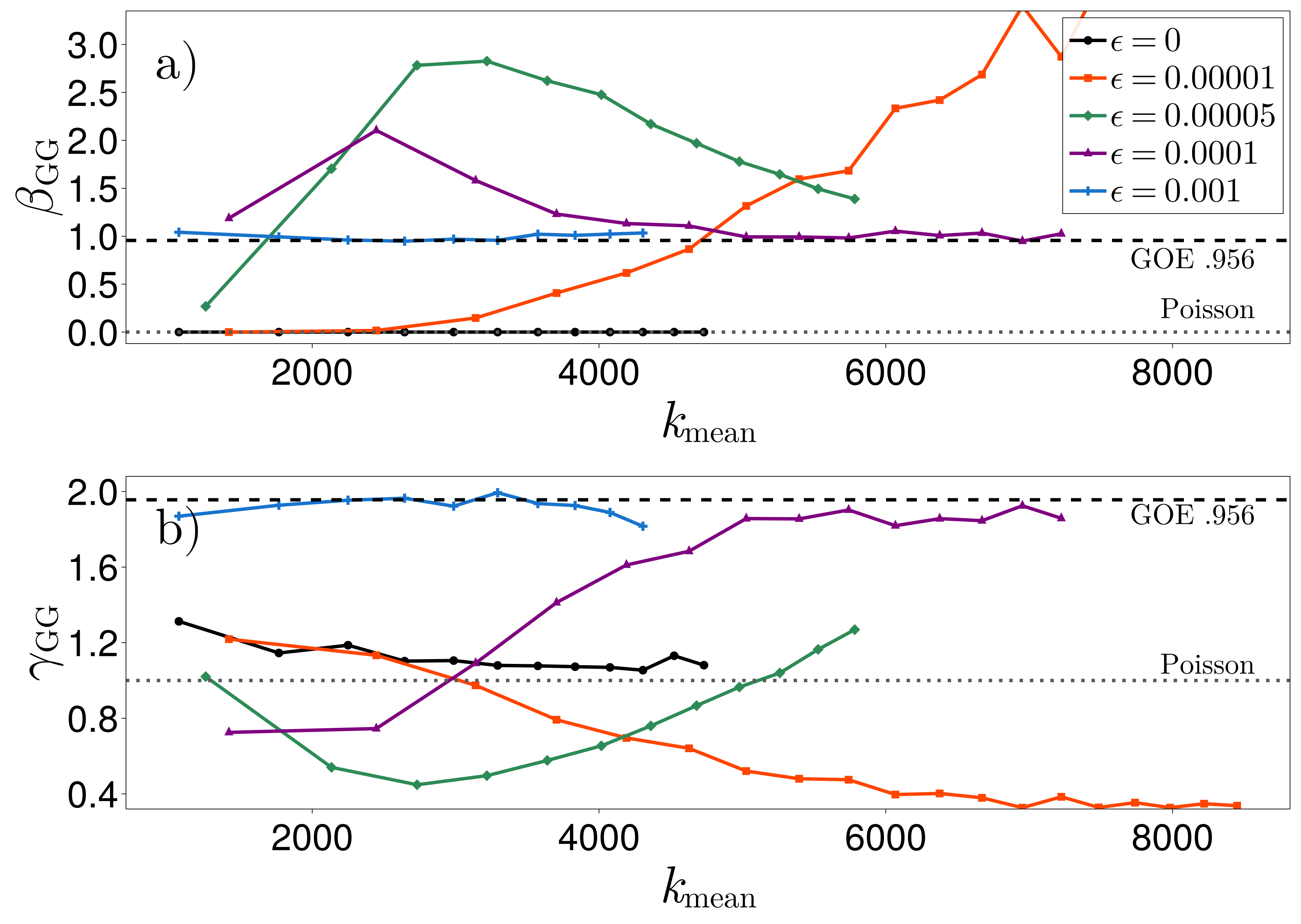}
    \caption{
    Change of the generalized-gamma fit parameters for the unfolded
    nearest-neighbour spacing distributions. The fits are performed on the
    conditional distribution for \(s\ge0.3\), so that the anomalous
    very-small-spacing region does not dominate the extracted parameters. Panel (a)
    shows the effective small- and intermediate-spacing parameter
    \(\beta_{\mathrm{GG}}\), while panel (b) shows the effective tail parameter
    \(\gamma_{\mathrm{GG}}\). The dotted gray lines mark the exact Poisson values
    \((\beta_{\mathrm{GG}},\gamma_{\mathrm{GG}})=(0,1)\). The dashed black lines
    mark the values corresponding to the best Brody approximation to the
    infinite-dimensional GOE spacing distribution,
    \(\beta_{\mathrm{GG}}\simeq0.956\) and
    \(\gamma_{\mathrm{GG}}\simeq1.956\). The non-monotone flow and the excursions
    outside the Poisson--GOE reference range show that the generalized-gamma form is
    not a universal crossover distribution for the perturbed hyperbolic billiards.
    Rather, it is used as a diagnostic fit: it reveals that the small/intermediate
    part of \(P(s)\) and the large-spacing tail relax towards GOE behaviour on
    different spectral scales.
    }
    \label{fig:parameter_flow_nnls}
\end{figure}

The fitted Brody and generalized-gamma parameters are shown in
Fig.~\ref{fig:parameter_flow_nnls}. These parameters should be interpreted as
effective shape diagnostics rather than as universal exponents. For
\(\epsilon=10^{-5}\), the Brody parameter remains small throughout the computed
range, whereas the generalized-gamma parameters indicate that the fitted bulk
and tail evolve differently. In particular, the effective tail parameter
\(\gamma_{\mathrm{GG}}\) remains far below the GOE reference value, reflecting
the persistence of long tails. For intermediate perturbations the parameter flow
is not monotone and can even leave the Poisson--GOE reference range. Such
excursions show that the fit is compensating for incompatible features of
\(P(s)\): developing level repulsion, a shifting peak, and a tail that remains
too heavy. Thus the generalized-gamma form should not be viewed as a crossover
law, but as a diagnostic showing that different parts of the spacing
distribution relax on different spectral scales.

The preceding fits show that, for the weakest perturbation
\(\epsilon=10^{-5}\), even the two-parameter generalized-gamma distribution is only a diagnostic description. Although it captures the bulk and the tail better than the one-parameter Brody distribution, it still does not fully reproduce the very-small-\(s\) structure and the simultaneous persistence of a heavy, nearly exponential tail. This suggests that the spectrum is not yet behaving as a single fully coupled GOE sequence. Instead, the data are consistent with a residual decomposition into several weakly coupled subspectra.

\begin{figure}
    \centering
    \includegraphics[width=\linewidth]{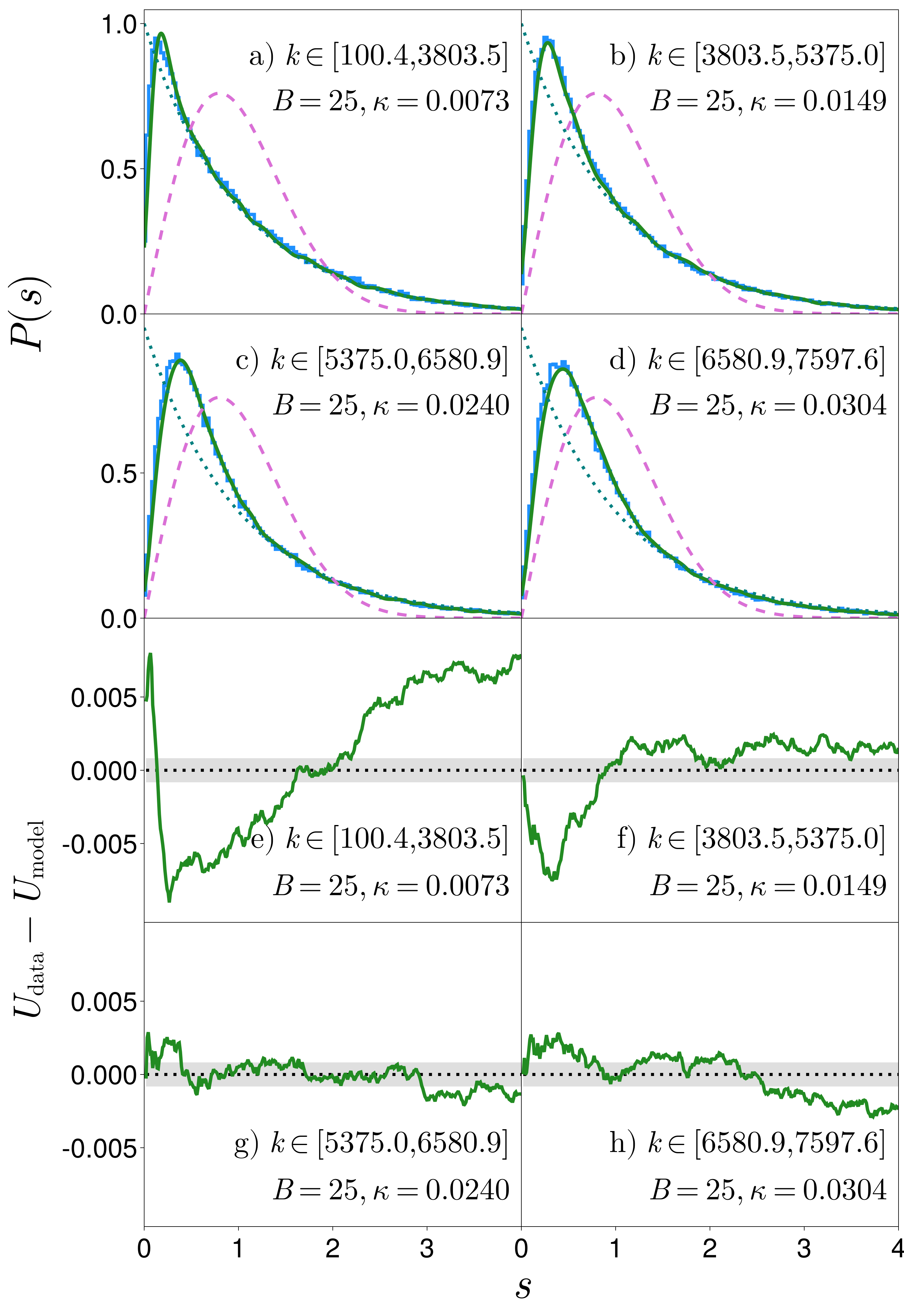}
    \caption{
    Block-GOE diagnostic for the nearest-neighbour spacing statistics of the weakly perturbed hyperbolic triangle with \(\epsilon=10^{-5}\) (number of GOE blocks is $B=25$ \eqref{eq:block_goe_model}). The four columns correspond to four consecutive spectral windows of \(1.5\times10^5\) eigenvalues each. Panels (a)--(d) show the spacing density \(P(s)\). The numerical data are shown as blue histograms, the fitted block-GOE model as the solid green curve, the Poisson distribution as the dotted teal curve, and the GOE Wigner surmise as the dashed magenta curve. Panels (e)--(h) show the corresponding deviation in the transformed cumulative representation, \(U_{\mathrm{data}}-U_{\mathrm{model}}\), where \(U(s)=(2/\pi)\arccos\sqrt{1-W(s)}\), with
    \(W(s)=\int_0^s P(s')\,ds'\). Following ~\cite{ProRob1994b}, this
    representation has approximately constant statistical uncertainty
    \(\delta U=1/(\pi\sqrt{N_s})\), where \(N_s\) is the number of spacings; the grey bands indicate \(\pm\delta U\). The fitted coupling parameter \(\kappa\) increases from \(\kappa=0.0073\) in the first window to \(\kappa=0.0304\) in the last window, indicating a slow drift away from the nearly decoupled block regime.
    }
    \label{fig:block_goe_eps_1e_minus_5}
\end{figure}

To quantify this picture we compare the \(\epsilon=10^{-5}\) data with a
finite-dimensional block-GOE crossover model. This is motivated by examples
from perturbed quantum maps, where additional symmetries produce spectra that
are superpositions of independent random-matrix subsequences
\cite{Cat_map_block_GOE_perturb}, and by deformed random-matrix models of
symmetry breaking in which initially uncoupled GOE blocks are coupled by random
off-diagonal matrix elements \cite{Defomed_ensembles_block_GOE}. We use
the following parametrization

\begin{equation}
H(\kappa)
=
\sqrt{1-\kappa^2}\,H_{\rm block}
+
\kappa H_{\rm GOE},
\label{eq:block_goe_model}
\end{equation}

\noindent where \(H_{\rm block}\) is block diagonal, with \(B\) independent and identical GOE blocks, and \(H_{\rm GOE}\) is a full GOE matrix of the same dimension as \(H_{\rm block}\). The parameter \(\kappa\in[0,1]\) controls the coupling between the blocks. For \(\kappa=0\), the spectrum is a superposition of \(B\) independent GOE subspectra and has an enhanced probability of small spacings and a heavier tail. For \(\kappa=1\), the model reduces to a single GOE spectrum. The fitted values of \(\kappa\) are small, indicating that the spectrum remains
far from the fully coupled GOE limit. At the same time, since $\kappa$ is nonzero, the model retains level repulsion in the strict \(s\to0\) limit,
\(P(s)\to0\). We use this model as an effective diagnostic of the residual decoupling of the spectrum.

The block-GOE comparison supports the interpretation suggested by the generalized-gamma fits (see Fig.\ref{fig:block_goe_eps_1e_minus_5}). For \(\epsilon=10^{-5}\), the spectrum is not simply moving along a universal Poisson--GOE interpolation.Instead, the spacing distributions retain features expected from a weakly
coupled superposition of GOE-like subsequences: enhanced near-zero spacing
weight and a heavy, nearly exponential tail coexist with a bulk that gradually
moves towards the GOE form. The fitted value of \(\kappa\) increases with \(k\), showing that the coupling between these components strengthens in the semiclassical direction, but the values remain small throughout the available range. This explains why the bulk of \(P(s)\) starts to deform while the tail remains close to exponential: different parts of the distribution are sensitive to different aspects of the residual decoupling. The first window in Fig.\ref{fig:block_goe_eps_1e_minus_5} (panel $a),e)$), where the residual
\(U_{\mathrm{data}}-U_{\mathrm{model}}\) is shown, exhibits a systematic
structure outside the statistical band. This shows that the simple block-GOE
ansatz does not fully describe the lowest-\(k\) regime.

Overall, the NNLS data shows a transition from the arithmetic
Poisson-like statistics of the Schmit triangle to generic GOE statistics under geometric perturbation. The transition is highly sensitive to the perturbation strength: \(\epsilon=10^{-5}\) remains close to the arithmetic regime over much of the accessible spectrum, \(\epsilon=10^{-4}\) flows steadily towards GOE, and \(\epsilon=10^{-3}\) is already essentially GOE. Across large ($\epsilon>10^{-5}$) perturbations, the generalized-gamma distribution provides a better effective global description than the Brody distribution, demonstrating that the crossover cannot be reduced to a single level-repulsion parameter.

\subsection{Spectral rigidity $\Delta_3(L)$}

We next consider the Dyson--Mehta spectral rigidity $\Delta_3(L)$ \cite{Mehta}, which probes
long-range correlations in the unfolded spectrum. For an unfolded sequence with
counting function $N(E)$, the local rigidity on an interval $[x,x+L]$ is defined
by the least-squares deviation of $N(E)$ from the best straight line,

\begin{equation}
\Delta_3(L;x) = \frac{1}{L} \min_{A,B} \int_x^{x+L} \left[N(E)-AE-B\right]^2\,dE .
\label{eq:delta3_local}
\end{equation}

\noindent The spectral rigidity is then obtained by averaging over the starting point $x$,

\begin{equation}
\Delta_3(L) = \left\langle \Delta_3(L;x)\right\rangle_x .
\label{eq:delta3_average}
\end{equation}

\noindent For a Poisson spectrum one has the linear growth

\begin{equation}
\Delta_3^{\mathrm{P}}(L)=\frac{L}{15},
\label{eq:delta3_poisson}
\end{equation}

\noindent whereas the GOE prediction grows only logarithmically for large $L$ \cite{Mehta}. Thus,
$\Delta_3(L)$ is particularly sensitive to the suppression of long-wavelength
fluctuations in the spectral staircase.

\begin{figure}[h!]
    \centering
    \includegraphics[width=\linewidth]{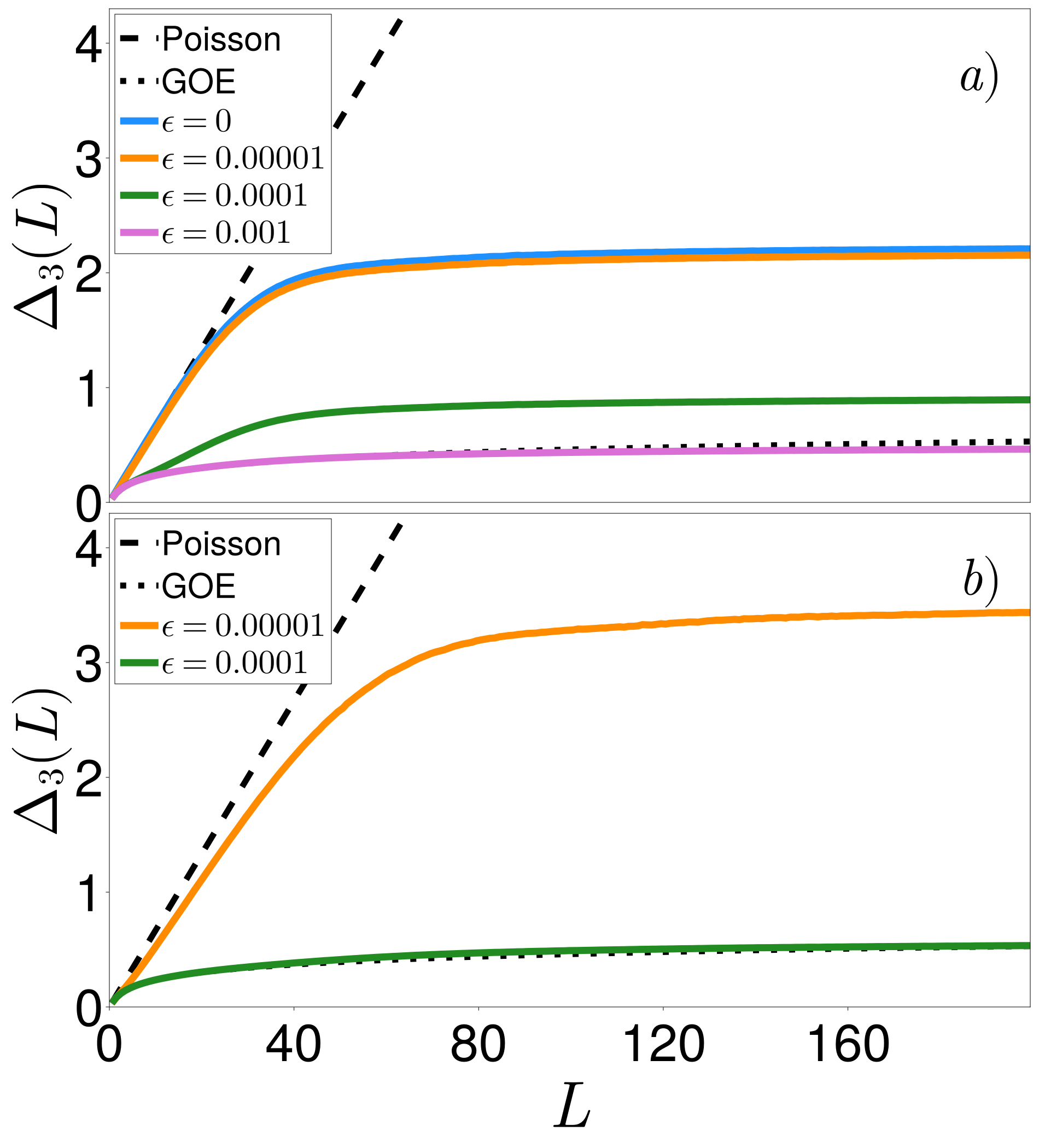}
   \caption{
    Spectral rigidity \(\Delta_3(L)\) for the Schmit triangle and its perturbations.
    In all curves shown here, \(\Delta_3(L)\) is computed from
    \(1.5\times10^5\) unfolded eigenvalues. Panel (a) uses the common level-index
    window \(n=60001,\ldots,210000\), which is available for all four perturbations
    \(\epsilon=0,10^{-5},10^{-4},10^{-3}\). Panel (b) uses the higher common
    level-index window \(n=433001,\ldots,583000\), which is available for the two
    smallest nonzero perturbations, \(\epsilon=10^{-5}\) and \(\epsilon=10^{-4}\).
    The dashed black line is the Poisson prediction \(\Delta_3(L)=L/15\), while the
    dotted black line is the GOE prediction. The unperturbed arithmetic triangle,
    \(\epsilon=0\), displays a strong saturation of \(\Delta_3(L)\), while the
    strongly perturbed case \(\epsilon=10^{-3}\) is close to GOE. The two weak
    perturbations show markedly different behaviour at high energies:
    \(\epsilon=10^{-4}\) approaches the GOE curve, whereas \(\epsilon=10^{-5}\)
    remains much closer to the rigid arithmetic behaviour over the accessible range.
    In panel (b), however, the \(\epsilon=10^{-5}\) curve has begun to bend towards
    the GOE prediction at small \(L\).
    }
    \label{fig:delta_3_last}
\end{figure}

Figure~\ref{fig:delta_3_last} shows $\Delta_3(L)$ computed from the last
$1.5\times10^5$ unfolded eigenvalues in common spectral windows. In panel (a),
the common upper cutoff is chosen so that all four spectra,
$\epsilon=0,10^{-5},10^{-4}$, and $10^{-3}$, contribute the same number of
levels. The unperturbed arithmetic triangle has a strongly saturated rigidity
curve, well below the Poisson line at large $L$. The smallest perturbation,
$\epsilon=10^{-5}$, remains close to this arithmetic behaviour over the common
range. By contrast, the larger perturbation $\epsilon=10^{-4}$ lies much closer
to GOE, while $\epsilon=10^{-3}$ is already essentially GOE-like on this scale.

Panel (b) compares only the two weakly perturbed systems, $\epsilon=10^{-5}$ and
$\epsilon=10^{-4}$, using their final common $1.5\times10^5$ levels. This probes
a higher part of the spectrum than panel (a), because more levels were computed
for these two cases. The difference between the two perturbations becomes even
clearer: the $\epsilon=10^{-4}$ curve approaches the GOE prediction, while the
$\epsilon=10^{-5}$ curve remains far more rigid and shows a delayed crossover.
Together with the nearest-neighbour spacing statistics and the sliding--window average gap
ratio, this confirms that the approach to GOE is strongly energy-dependent and
becomes extremely slow for very small perturbations of the arithmetic triangle.

\subsection{Spectral form factor (SFF)}

As a complementary long-range statistic we compute the spectral form factor. The
levels are first unfolded using the Weyl law \eqref{eq:weyl_k}. For a window of
\(M\) consecutive unfolded levels \(\epsilon_j\), we define the finite-window
spectral form factor by

\begin{equation}
K_M(\tau)
=
\frac{1}{M}
\left|
\sum_{j=1}^{M}
\exp\left(2\pi i\tau\epsilon_j\right)
\right|^2 .
\label{eq:sff_definition}
\end{equation}

\noindent The SFF of a single finite spectrum is not self-averaging and is
therefore strongly oscillatory. For random-matrix ensembles or disordered systems
one can average over realizations. For clean billiards, where only one spectrum is
available, the standard substitute is a spectral-window average together with a
smoothing in \(\tau\) \cite{Lozej_Bogomolny_triangle_gamma}. We therefore average
\eqref{eq:sff_definition} over overlapping windows of \(M=800\) consecutive
levels, giving a window-averaged form factor \(\overline K_M(\tau)\). To suppress
the remaining finite-window oscillations, we then apply a Gaussian convolution in
\(\tau\). The plotted quantity is

\begin{equation}
K_\sigma(\tau)
=
\frac{
\displaystyle
\int_0^\infty
\exp\left[-\frac{(\tau-\tau')^2}{2\sigma^2}\right]
\overline K_M(\tau')\,d\tau'
}{
\displaystyle
\int_0^\infty
\exp\left[-\frac{(\tau-\tau')^2}{2\sigma^2}\right]d\tau'
}.
\label{eq:sff_gaussian_smoothing}
\end{equation}

\begin{figure}[h!]
    \centering
    \includegraphics[width=\linewidth]{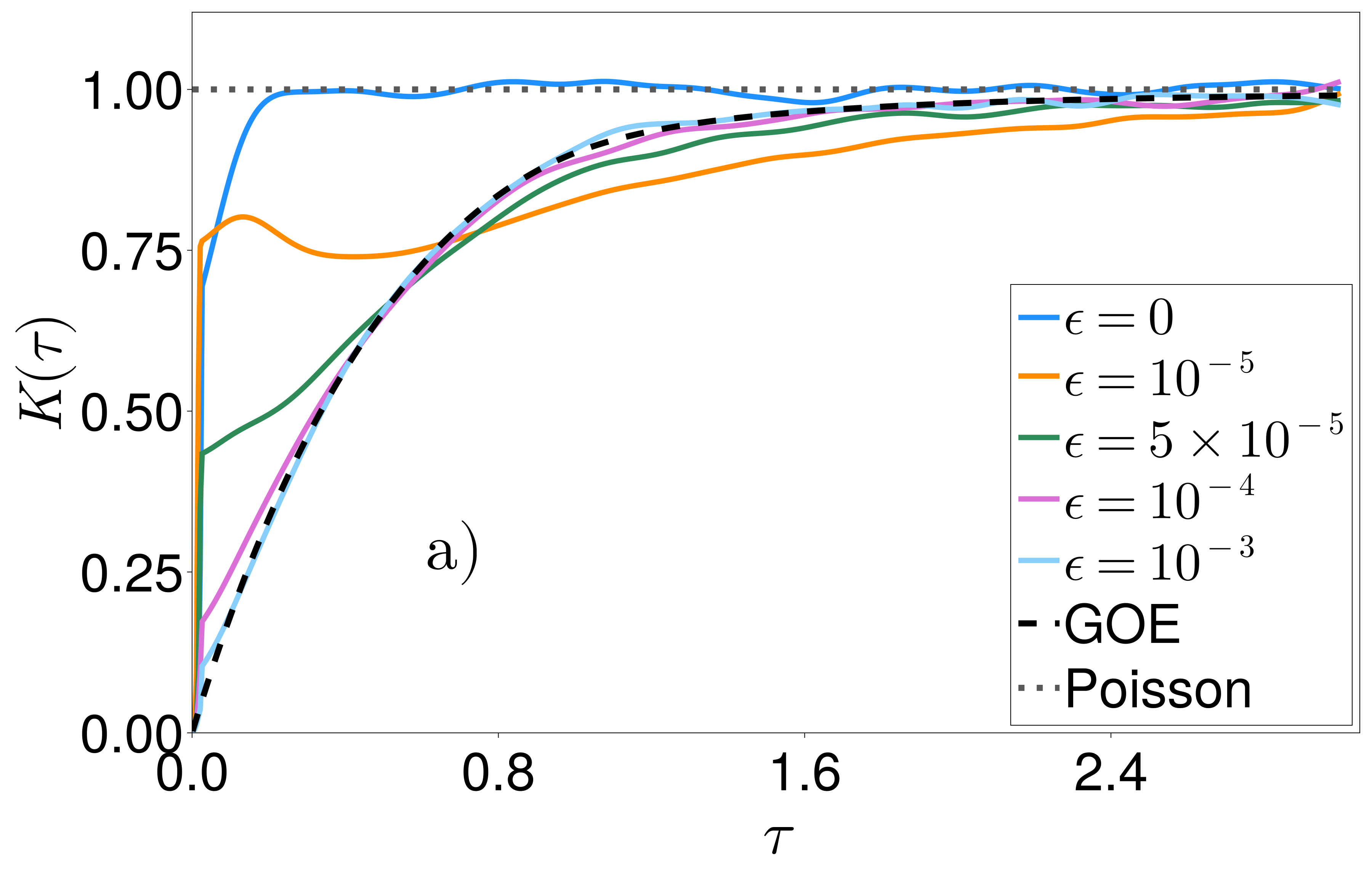}

    \vspace{0.2cm}

    \includegraphics[width=\linewidth]{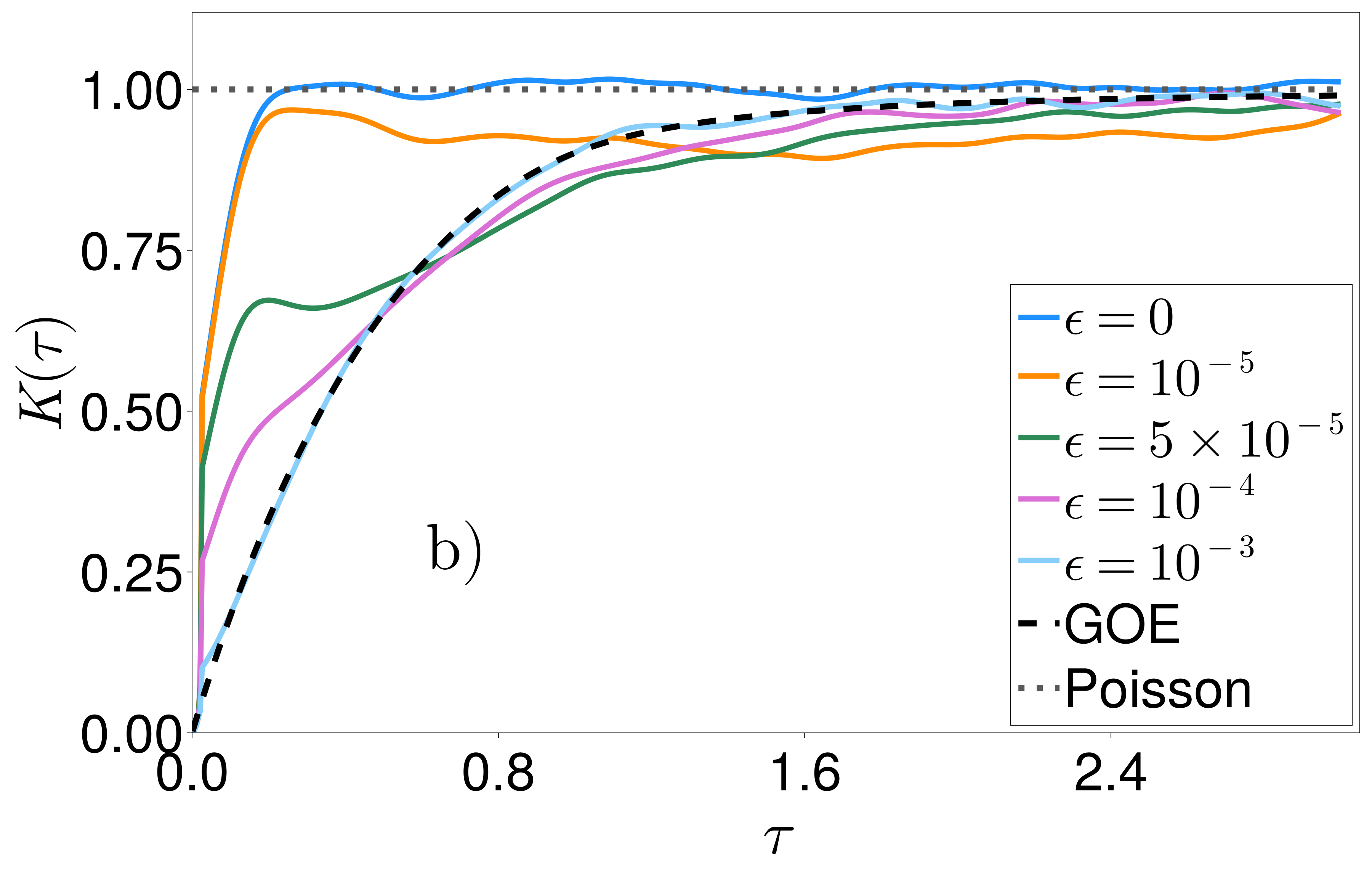}
    \caption{
    Spectral form factor \(K(\tau)\) for the Schmit triangle and its
    perturbations. The spectra are unfolded with the Weyl law
    \eqref{eq:weyl_k}. The finite-window form factor is computed in overlapping
    windows of \(M=800\) consecutive unfolded levels and then smoothed by a Gaussian
    in \(\tau\) with width \(\sigma=0.08\), as defined in
    \eqref{eq:sff_gaussian_smoothing}. The dashed black curve is the
    infinite-dimensional GOE prediction \eqref{eq:sff_goe}, while the dotted gray
    line is the Poisson prediction \(K_{\rm P}(\tau)=1\). Panel (a) uses the full
    available spectrum for each perturbation: \(210000\) eigenvalues for
    \(\epsilon=0\), \(782000\) for \(\epsilon=10^{-5}\), \(583000\) for
    \(\epsilon=10^{-4}\), and \(210000\) for \(\epsilon=10^{-3}\). Panel (b) uses
    the common level-index window \(n=1,\ldots,150000\) for each perturbation,
    giving a comparison at equal spectral depth. In the full spectra, the larger
    perturbations \(\epsilon=10^{-4}\) and \(\epsilon=10^{-3}\) approach the GOE
    curve, whereas \(\epsilon=10^{-5}\) remains visibly closer to the arithmetic
    regime.
    }
    \label{fig:sff}
\end{figure}

In the computations we used \(\sigma=0.08\) for all spectra. The value
\(\tau=0\) itself was not used: for a finite spectrum the diagonal contribution
makes \(K_M(0)=M\), so the small-\(\tau\) limit is meaningful only after the
large-spectrum limit has been taken. In practice we start the calculation at
\(\tau_{\min}=1/M\), and the first few values near \(\tau=0\) are left unsmoothed
in order not to distort the behaviour close to the origin. We therefore interpret
Fig.~\ref{fig:sff} as a comparison of the smoothed finite-spectrum form factor
with the universal Poisson and GOE limits, rather than as a pointwise estimate of
the unsmoothed \(K_M(\tau)\). For comparison we use the Poisson prediction

\begin{equation}
K_{\rm P}(\tau)=1,
\label{eq:sff_poisson}
\end{equation}

\noindent and for GOE

\begin{equation}
K_{\rm GOE}(\tau)
=
\begin{cases}
2\tau-\tau\log(1+2\tau),
& 0\leq \tau\leq 1, \\[1mm]
2-\tau\log\left(\dfrac{2\tau+1}{2\tau-1}\right),
& \tau\geq 1 .
\end{cases}
\label{eq:sff_goe}
\end{equation}

The results are shown in Fig.~\ref{fig:sff}. Panel (a), computed from the full
available spectrum for each perturbation, shows that the larger perturbations
approach the GOE prediction, while the unperturbed arithmetic triangle remains
close to the Poisson plateau. The weakest perturbation, \(\epsilon=10^{-5}\),
also remains much closer to Poisson than to GOE over the available range,
although a visible deviation from the Poisson plateau is already present.

Panel (b) compares the spectra at the same level count, using only the first
\(1.5\times10^5\) eigenvalues for each perturbation. This removes the effect of
different total spectral ranges and shows the perturbation dependence more
directly. At the same spectral depth, smaller perturbations remain farther from
the GOE curve, whereas \(\epsilon=10^{-3}\) is already close to GOE. The
comparison between panels (a) and (b) confirms that the approach to GOE is both
perturbation-dependent and energy-dependent: increasing the perturbation or
moving higher in the spectrum both drive the form factor towards the GOE
prediction.

\section{Entropy localization measures}
\label{localization_measures}

\subsection{Poincar\'e--Husimi (PH) function}
\label{subsec:poincare_husimi}

To study the phase-space structure of the eigenfunctions, we use a
Poincar\'e--Husimi (PH) representation on the billiard boundary. For a billiard,
the boundary provides a natural Poincar\'e section. We parametrize it by the
hyperbolic arclength coordinate \(q \in [0,L)\) along the boundary and by the
tangential momentum

\begin{equation}
    p = \sin \alpha \in [-1,1],
    \label{eq:ph_p_sin_chi}
\end{equation}

\noindent where \(\alpha\) is the angle of incidence measured from the inward hyperbolic
normal. Thus \(p=0\) corresponds to normal incidence, while \(|p|=1\) corresponds
to grazing motion.

For Dirichlet eigenfunctions, the boundary values of the eigenfunction vanish,
and the relevant boundary data are instead the normal derivatives

\begin{equation}
    u_j(q) = \partial_n \psi_j(q),
    \label{eq:ph_boundary_function}
\end{equation}

\noindent where \(\partial_n\) denotes the unit outward hyperbolic normal derivative. These boundary
functions determine the corresponding eigenfunctions in the interior and are
therefore a natural reduced representation of the quantum state. Following the
boundary coherent-state construction of PH functions for billiards
\cite{TV1995,Backer2003}, we project \(u_j\) onto Gaussian coherent states localized at points \((q,p)\) of
the boundary section.

In the present hyperbolic setting we use the same local construction, with
Euclidean arclength and normal derivatives replaced by their hyperbolic
counterparts. The coherent state centered at \((q,p)\) is taken to be the
periodized Gaussian

\begin{equation}
    c_{q,p}^{(k)}(s)
    =
    \left(\frac{k}{\pi}\right)^{1/4}
    \sum_{m\in\mathbb{Z}}
    \exp\left[
        i k p (s-q+mL)
        -\frac{k}{2}(s-q+mL)^2
    \right],
    \label{eq:ph_boundary_coherent_state}
\end{equation}

\noindent where \(s\) is hyperbolic arclength along the boundary and \(L\) is the total
hyperbolic boundary length. The width of the coherent state is therefore of
order \(k^{-1/2}\), as expected semiclassically. The corresponding PH density is
defined by

\begin{equation}
    H_j(q,p)
    =
    \frac{1}{2\pi k_j}
    \left|
        \int_{\partial\Omega}
        {c_{q,p}^{(k_j)}(s)}\,u_j(s)\,ds_{H}
    \right|^2 .
    \label{eq:ph_density}
\end{equation}

\noindent After normalization over the boundary section, \(H_j(q,p)\) is interpreted as a
probability density on the classical Poincar\'e section.

\subsection{Beta distribution}
\label{subsec:beta_distribution}

To quantify phase-space localization statistically, we use the entropy of the normalized PH function. For each eigenstate, the PH function is discretized on a uniform $(q,p)$ grid and normalized as

\begin{equation}
\sum_{q,p} H_j(q,p)=1 .
\label{eq:ph_grid_normalization}
\end{equation}

\noindent The corresponding entropy is

\begin{equation}
S_j
=
-\sum_{q,p} H_j(q,p)\log H_j(q,p).
\label{eq:ph_entropy}
\end{equation}

\noindent From this we define the normalized entropy localization measure \cite{BatRob2013A}

\begin{equation}
A_j
=
\frac{\exp(S_j)}{N_{\mathrm{eff}}},
\label{eq:entropy_localization_measure}
\end{equation}

\noindent where $N_{\mathrm{eff}}$ is the number of effectively accessible phase-space cells. By construction, $A_j$ is small for strongly localized PH functions, such as scarred or corner-localized states, and larger for eigenstates distributed over a substantial part of the Poincar\'e section. In the present normalization, fully extended chaotic states concentrate near $A\simeq 0.6$–$0.65$, rather than near $A=1$, due to the non-uniform mean structure of the PH representation and the finite coherent-state resolution.

We study the distribution $P(A)$ in spectral windows. Empirically, these distributions are well described by a beta distribution on a finite interval $0<A<A_0$ \cite{BLR2020},

\begin{equation}
P(A;\alpha,\beta)
=
CA^{\alpha-1}(A_0-A)^{\beta-1},
\quad
0<A<A_0 ,
\label{eq:beta_distribution_A}
\end{equation}

with normalization

\begin{equation}
C^{-1}
=
A_0^{\alpha+\beta-1}B(\alpha,\beta),
\quad
B(\alpha,\beta)
=
\int_0^1 t^{\alpha-1}(1-t)^{\beta-1}dt .
\label{eq:beta_distribution_normalization}
\end{equation}

\noindent The shape parameters $\alpha$ and $\beta$ encode the asymmetry and concentration of the localization-measure distribution. In particular, the standard deviation of the fitted distribution,

\begin{equation}
\sigma_{\beta}
=
A_0
\sqrt{
\frac{\alpha\beta}
{(\alpha+\beta)^2(\alpha+\beta+1)}
},
\label{eq:beta_distribution_std}
\end{equation}

\noindent provides a compact measure of the width of $P(A)$. A decrease of $\sigma_{\beta}$ with increasing $k$ therefore signals the suppression of phase-space localization fluctuations in the semiclassical limit.

\begin{figure}[h!]
\centering
\vspace{-0.1cm}
\textbf{$\epsilon=0$}

\includegraphics[width=0.9\linewidth]{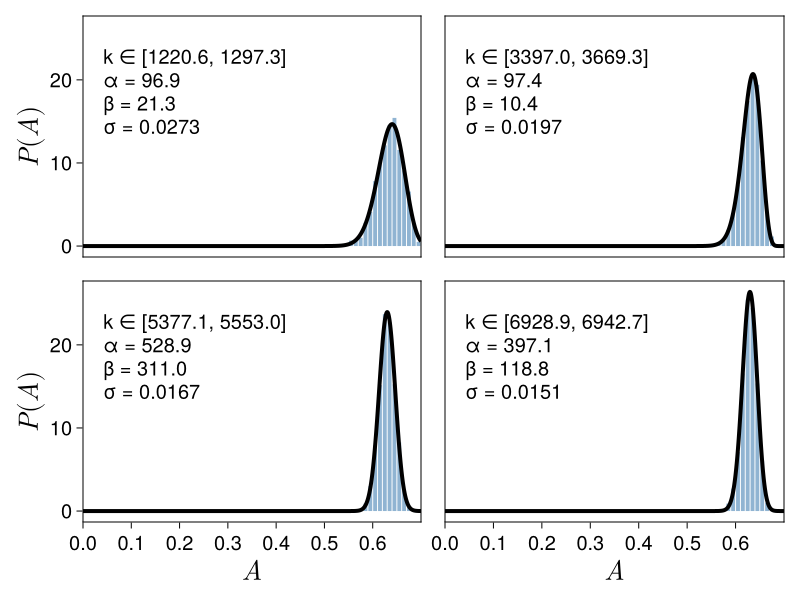}
\vspace{-0.1cm}
\noindent\rule{0.95\linewidth}{0.7pt}
\textbf{$\epsilon=0.0001$}
\includegraphics[width=0.9\linewidth]{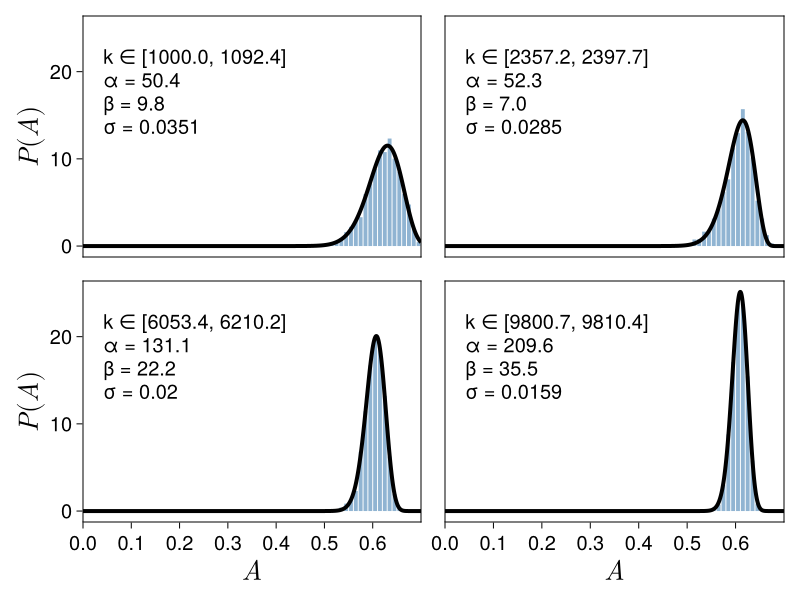}
\vspace{-0.1cm}
\noindent\rule{0.95\linewidth}{0.7pt}
\textbf{$\epsilon=0.001$}
\includegraphics[width=0.9\linewidth]{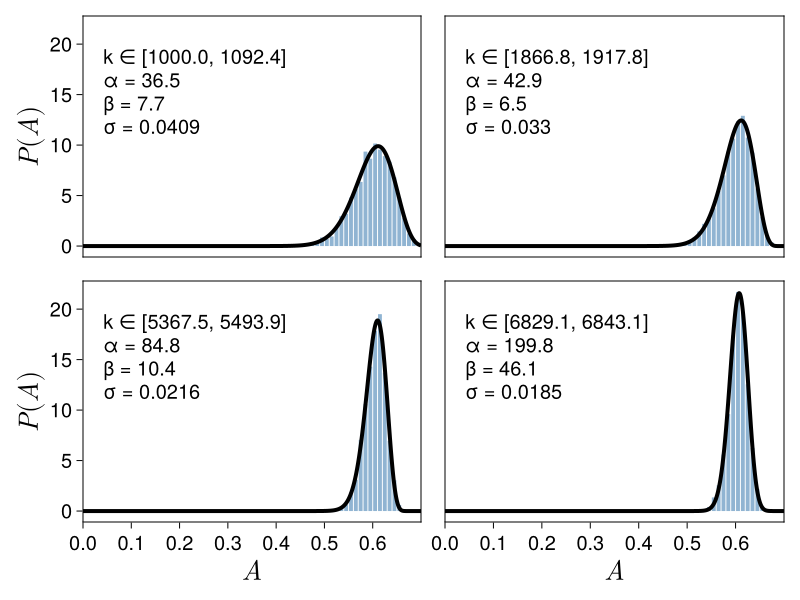}
\caption{
    Distribution $P(A)$ of the phase-space entropy localization measure $A$ for the hyperbolic triangle billiards, fitted by beta distributions. The three blocks correspond, from top to bottom, to $\epsilon=0$, $\epsilon=10^{-4}$, and $\epsilon=10^{-3}$. Each panel shows one spectral window, with the empirical histogram in blue and the fitted beta distribution in black. The fitted parameters $\alpha$, $\beta$, and standard deviation $\sigma$ are displayed in each panel. The narrowing with increasing $k$ indicates a decreasing variance of the entropy localization measure.
}
\label{fig:betas}

\end{figure}

Representative distributions are shown in Fig.~\ref{fig:betas}. For all three perturbations, the distributions narrow as $k$ increases, showing that the ensemble of PH functions becomes increasingly self-averaging. This is consistent with the expected semiclassical suppression of exceptional localized states.

\begin{figure}
\centering
\includegraphics[width=\linewidth]{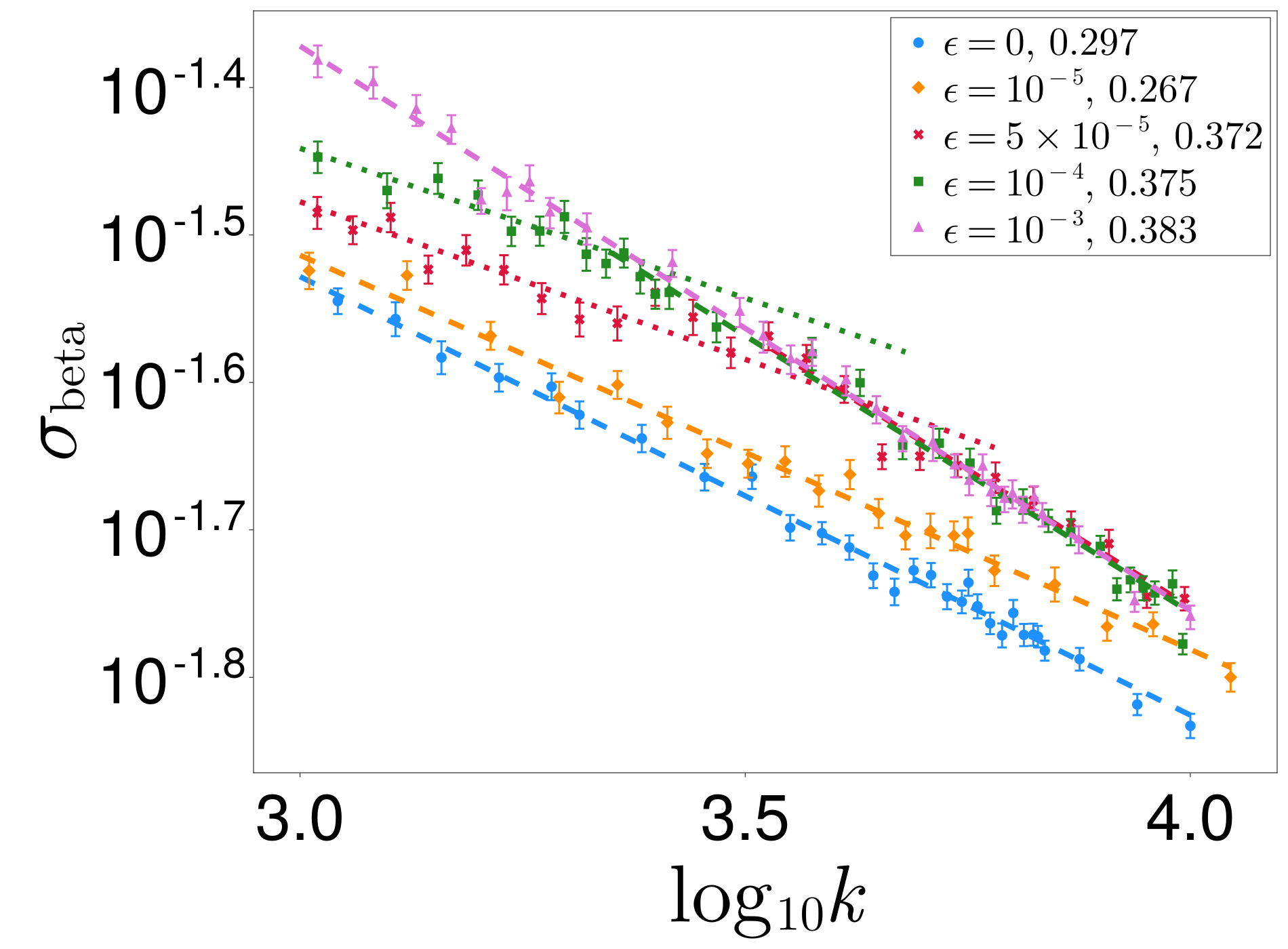}
\caption{
    Power-law decay of phase-space localization fluctuations across the Poisson--GOE crossover. The standard deviation \(\sigma_{\beta}\) of the beta-distribution fit to the entropy localization measure is shown as a function of \(\log_{10}k\) for several perturbations \(\epsilon\). Symbols
    denote spectral-window estimates, with error bars from the beta fits, and the lines show power-law fits \(\sigma_{\beta}\sim k^{-\eta}\); the fitted
    exponents \(\eta\) are given in the legend. Each point contains $N=2000$ levels for which \eqref{eq:beta_distribution_A} was computed. The curves are separated in the pre-crossover regime, reflecting the different rates at which the perturbed systems approach GOE spectral statistics. In particular, the crossover occurs
    near \(k\simeq 10^{3.5}\) for \(\epsilon=10^{-4}\) and near
    \(k\simeq 10^{3.65}\) for \(\epsilon=5\times10^{-5}\), consistent with the crossover scales extracted independently from the Weyl-law deviation analysis in Fig.~\ref{fig:weyl}. Beyond these scales, the perturbations that reach the
    GOE regime collapse onto a common decay line, indicating that the remaining phase-space localization fluctuations are governed by the same semiclassical mechanism. For \(\epsilon=10^{-5}\), the crossover lies beyond the directly
    accessible range; extrapolation places it near \(k\simeq10^{4.1}\), corresponding to approximately the \(1.65\times10^6\)-th level.
}
\label{fig:main_power_law}
\end{figure}

Different parts of a finite spectrum can exhibit different degrees of
eigenfunction localization. Such dynamical localization is expected to weaken in
the semiclassical limit, and its suppression provides a useful eigenfunction
signature of the approach to quantum ergodicity. In strongly localized regimes,
however, the distribution of localization measures may cease to be well
described by a Beta law, because isolated families of strongly localized states
produce non-universal tails or secondary structures \cite{batistic_robnik_localization_eigenstates}. The present data are
therefore favourable: the PH entropy distributions remain broadly delocalized
and are well fitted by Beta distributions throughout the analyzed windows. This
allows the fitted Beta width to be used as a robust measure of localization
fluctuations across the arithmetic-to-GOE crossover. 
The dependence of $\sigma_{\beta}$ on $k$ is shown in Fig.~\ref{fig:main_power_law}. We fit the windowed values by a power law,

\begin{equation}
\sigma_{\beta}(k)
\sim
k^{-\eta},
\label{eq:sigma_beta_power_law}
\end{equation}

\noindent where $\eta$ is extracted from a linear fit on the log–log scale. The resulting exponents are displayed in the legend of Fig.~\ref{fig:main_power_law}.

The key observation is that the apparent decay exponent depends on whether the
spectral window lies before or after the Poisson--GOE crossover. Before the
crossover, the localization statistics retain memory of the arithmetic or
near-arithmetic regime, and the curves remain separated. Once a perturbation
reaches the GOE spectral regime, however, its \(\sigma_{\beta}\) curve falls onto
the same decay line as the other GOE-limited perturbations. Thus the power-law
decay of the beta width separates two effects: the perturbation-dependent scale
at which GOE statistics emerge, and a common post-crossover suppression of
phase-space localization fluctuations.

For the unperturbed arithmetic case \(\epsilon=0\), the spectral statistics
remains non-GOE and the decay of \(\sigma_{\beta}\) does not collapse onto the
a common line (see Fig.\ref{fig:main_power_law}). The weak perturbation \(\epsilon=10^{-5}\) also remains
pre-asymptotic over the computed range, consistent with the extrapolated
crossover scale lying at substantially larger \(k\). In contrast,
\(\epsilon=5\times10^{-5}\), \(\epsilon=10^{-4}\), and \(\epsilon=10^{-3}\)
reach the GOE regime within the available spectral data, after which their
localization-fluctuation widths follow the same power--law decay.

The resulting post-crossover exponents are of order
\(\eta\simeq0.37\)--\(0.4\) for the perturbations that reach the GOE limit.
This is close to the exponents previously obtained in the fully chaotic
\(C_3\) billiard in the Euclidean metric \cite{c3_arxiv_orel}. In that system, the prefactor
of the power-law decay depended systematically on the average
maximal Lyapunov exponent of the classical dynamics. In the present hyperbolic
triangles, by contrast, the geodesic flow has a common average maximal Lyapunov exponent,
\(\lambda=1\), for all perturbations. This provides a natural explanation for
why, after the Poisson--GOE crossover, the curves collapse onto the same line:
once the arithmetic or near-arithmetic transient has disappeared, the remaining
localization fluctuations are governed by the same ergodic hyperbolic dynamics.

\section{Conclusion}
\label{conclusion}

We have studied the spectral and eigenfunction properties of a family of
hyperbolic triangle billiards obtained by perturbing the arithmetic Schmit
triangle. The perturbation preserves the area of the billiard while breaking the
exact arithmetic structure responsible for the non-generic spectral statistics
of the unperturbed system. This provides a controlled setting in which to study
how generic quantum chaos emerges from an arithmetic hyperbolic billiard in the
semiclassical limit.

The spectral statistics show a clear perturbation-dependent crossover from the
arithmetic, Poisson-like regime to GOE statistics. For the unperturbed triangle,
the nearest-neighbour spacing distribution, sliding--window average gap ratio, spectral rigidity,
and spectral form factor remain close to the arithmetic behaviour throughout the
computed range. For the largest perturbation, \(\epsilon=10^{-3}\), the transition
to GOE is already essentially complete in the lowest spectral windows studied.
The intermediate perturbations display a scale-dependent crossover: for
\(\epsilon=10^{-4}\) the GOE regime is reached within the available spectrum,
whereas for \(\epsilon=10^{-5}\) the spectrum remains largely arithmetic-like over
the computed range.

A central point of the analysis is that the crossover is not well described by a
single Brody parameter. The spacing distributions show that the small-spacing
behaviour and the large-spacing tail evolve on different scales. The
generalized-gamma distribution gives a better global description because it
separates the level-repulsion exponent from the tail-decay exponent. However, it
is not an exact description of the crossover distribution: in particular, it
captures the bulk and the tail only qualitatively and does not reproduce the
very-small-\(s\) behaviour accurately. This shows that the transition from
arithmetic to GOE statistics is not a one-parameter interpolation, but rather a
gradual deformation of the full spacing distribution. This behaviour is also
reflected in the spectral rigidity and spectral form factor, which show that
long-range correlations approach GOE only at sufficiently high energies and
sufficiently large perturbations.

For the weakest perturbation, \(\epsilon=10^{-5}\), the block-GOE comparison
provides an additional diagnostic of this slow crossover. The fitted model is
consistent with a long-lived regime in which the spectrum is better viewed as
several weakly coupled GOE-like subspectra than as a single fully coupled GOE
sequence. The fitted coupling parameter increases with \(k\), but remains small
over the accessible spectral range. This supports the interpretation that the
bulk of the spacing distribution can begin to deform before the large-\(s\) tail
has fully crossed over to GOE-like behaviour.

We have also examined the eigenfunctions through their Poincar\'e--Husimi
representations on the boundary phase space. The entropy-based localization
measure provides a compact way of quantifying the spread of the PH functions.
Its distribution in spectral windows is well described by Beta distributions,
whose fitted width decreases as a power--law with wavenumber. Before the spectral
crossover, the localization statistics remain separated and retain memory of the
arithmetic or near-arithmetic regime. After the GOE regime is reached, however,
the curves for the perturbations with GOE limits collapse onto a common power
law,

\begin{equation}
    \sigma_{\mathrm{Beta}}(k)\sim k^{-\eta},
    \qquad
    \eta\simeq 0.37\text{--}0.4 .
    \label{eq:conclusion_beta_width_decay}
\end{equation}

\noindent Here \(\sigma_{\mathrm{Beta}}\) denotes the standard deviation of the fitted Beta
distribution. The exponent is close to the one previously found for fully
chaotic \(C_3\)-symmetric Euclidean billiards. The agreement suggests that, once
the arithmetic transient has disappeared, the decay of localization fluctuations
is governed by a generic ergodic self-averaging mechanism rather than by the
specific details of the perturbation.

The numerical results therefore give a consistent picture at both the spectral
and eigenfunction levels. Breaking arithmeticity does not immediately produce
GOE statistics at all energies; instead, the system passes through a long
perturbation-dependent crossover regime. The smaller the perturbation, the
farther this crossover is pushed into the large-\(k\) part of the spectrum. Once
the crossover is complete, both the spectral statistics and the phase-space
localization fluctuations behave as in generic time-reversal-invariant chaotic
systems. These results demonstrate that large consecutive spectra and
eigenfunction ensembles are essential for resolving the arithmetic-to-GOE
transition in hyperbolic quantum billiards.

\subsection{Acknowledgements}

I thank Prof. Dr. Marko Robnik for helpful discussions and careful reading of the manuscript and Dr. Hua Yan for useful comments and inspiring discussions. This work was supported by the Slovenian Research and Innovation Agency (ARIS) under grants J1-4387 and P1-0306, and made use of the HPC VEGA supercomputer system under project S24O02-01.

\subsection{Data Availability}

All data made for this study is found and can be calculated using the library \cite{QuantumBilliards_Orel} (GitHub) living on the \texttt{hyperbolic} branch. The relevant files for the calculation of energy levels and Poincar\'e--Husimi function are found in the \texttt{examples/hyperbolic} folder.

\bibliography{main}

\appendix

\subsection{Analytic logarithmic split of the hyperbolic double-layer kernel}
\label{app:hyp_kress_split}

The Nyström discretization \cite{kress_book} requires the logarithmic singularity of the
double-layer kernel to be separated analytically. We use the computational
periodic variable \(\sigma\) for the Kress logarithm \cite{martensen1963,kussmaul1969,kress} and write the physical
boundary point as \(x=x(t(\sigma))\). For a smooth boundary, \(t(\sigma)=\sigma\);
for a corner-graded boundary, \(t(\sigma)\) is the grading map defined in
Appendix.~\ref{app:nystrom_kress}. The goal is to write the jump-scaled kernel as

\begin{equation}
\begin{aligned}
K(\sigma,\eta)
&=
L_1(\sigma,\eta)\,
\mathcal R(\sigma,\eta)
+
L_2(\sigma,\eta),
\\[2mm]
\mathcal R(\sigma,\eta)
&=
\log\left(
4\sin^2\frac{\sigma-\eta}{2}
\right),
\end{aligned}
\label{eq:kress_split_goal}
\end{equation}

\noindent where \(L_1\) and \(L_2\) have smooth diagonal limits on each smooth boundary
arc. The only difference from the Euclidean Helmholtz case is the conformal
correction in the diagonal value of \(L_2\).

Let \(d_H\) denote the geodesic distance induced by the hyperbolic metric
\(g_H\) \eqref{eq:poincare_metric}. In the Poincar\'e disk, the hyperbolic distance
\(\chi(x,y)=d_H(x,y)\) is determined by \cite{aurich_equation_wavefunction_hyperbolic}

\begin{equation}
\cosh\chi(x,y)
=
1+
\frac{2|x-y|^2}
{(1-|x|^2)(1-|y|^2)} .
\label{eq:hyper_distance}
\end{equation}

\noindent We will need the derivative of \(\chi\) with respect to the source point
\(y\) in the Euclidean outward normal direction \(n_y\). Differentiating
\eqref{eq:hyper_distance} gives
\begin{equation}
\sinh\chi(x,y)\,\partial_{n_y}\chi(x,y)
=
\partial_{n_y}
\left[
1+
\frac{2|x-y|^2}
{(1-|x|^2)(1-|y|^2)}
\right].
\label{eq:hyper_distance_derivative_step}
\end{equation}

\noindent Since \(x\) is fixed in this differentiation, this yields

\begin{equation}
\partial_{n_y}\chi(x,y)
=
\frac{4\,n_y\cdot
\left(
y-x+
\frac{|x-y|^2}{1-|y|^2}\,y
\right)}
{(1-|x|^2)(1-|y|^2)\sinh\chi(x,y)} .
\label{eq:hyper_distance_normal}
\end{equation}

The diagonal behaviour of this expression is important for the Kress split.
Let \(x\) and \(y\) approach one another along the same smooth boundary arc.
Then the chord \(y-x\) is tangent to the boundary to leading order, and hence

\[
n_y\cdot(y-x)=O(|x-y|^2).
\]

\noindent The second term in the numerator of \eqref{eq:hyper_distance_normal} is also
\(O(|x-y|^2)\). Therefore the whole numerator is \(O(|x-y|^2)\). On the other
hand, \(\chi(x,y)=O(|x-y|)\) and
\(\sinh\chi=\chi+O(\chi^3)\). Hence

\begin{equation}
\partial_{n_y}\chi(x,y)
=
O(\chi(x,y)),
\quad x\to y.
\label{eq:hyper_distance_normal_order}
\end{equation}

\noindent The Green function is then \cite{Aurich_Green_def}

\begin{equation}
G_k(x,y)
=
\frac{1}{2\pi}
Q_\nu(\cosh\chi(x,y)),
\quad
\nu=-\frac12+ik,
\label{eq:green_legendre}
\end{equation}

\noindent where $Q_\nu$ is the associated Legendre function of the second kind. Primes denote differentiation with respect to the hyperbolic distance \(\chi\), not with respect to the Legendre argument. Thus

\begin{equation}
Q_\nu'(\cosh\chi)
=
\frac{d}{d\chi}Q_\nu(\cosh\chi),
\quad
P_\nu'(\cosh\chi)
=
\frac{d}{d\chi}P_\nu(\cosh\chi).
\label{eq:prime_definition}
\end{equation}

\noindent Locally, as \(\chi\to0\) \cite{NIST:DLMF},

\begin{equation}
Q_\nu(\cosh\chi)
=
-P_\nu(\cosh\chi)\log\frac{\chi}{2}
+
B_\nu(\chi),
\label{eq:Q_split}
\end{equation}

\noindent where \(B_\nu\) is smooth. Hence

\begin{equation}
Q_\nu'(\cosh\chi)
=
-P_\nu'(\cosh\chi)\log\frac{\chi}{2}
-
\frac{P_\nu(\cosh\chi)}{\chi}
+
B_\nu'(\chi).
\label{eq:Qprime_split}
\end{equation}

\noindent Equation~\eqref{eq:Q_split} is used only to identify the singular structure
near the diagonal. Away from the diagonal, the decomposition remains exact: the
coefficient of the logarithm is extended using the exact smooth function
\(P_\nu(\cosh\chi)\), and the remainder is defined by subtraction from the exact
kernel.

The jump-scaled source-normal kernel used in the discretization is

\begin{equation}
K(x,y)
:=
2\partial_{n_y}G_k(x,y)
=
\frac{1}{\pi}
Q_\nu'(\cosh\chi(x,y))\,\partial_{n_y}\chi(x,y).
\label{eq:jump_scaled_kernel}
\end{equation}

\noindent Substituting ~\eqref{eq:Qprime_split} gives

\begin{equation}
\begin{aligned}
K(x,y)
={}&
-\frac{1}{\pi}
P_\nu'(\cosh\chi)\partial_{n_y}\chi
\log\frac{\chi}{2} \\
&-\frac{1}{\pi}
\frac{P_\nu(\cosh\chi)}{\chi}
\partial_{n_y}\chi
+
\frac{1}{\pi}
B_\nu'(\chi)\partial_{n_y}\chi .
\end{aligned}
\label{eq:kernel_decomposed}
\end{equation}

\noindent The first term contains the explicit logarithm. The inverse-\(\chi\) term has a
finite boundary limit on a smooth boundary and determines the diagonal value of
\(L_2\).

We now replace the local logarithm by the periodic Kress logarithm. For the
computational parameters \(\sigma,\eta\in[0,2\pi)\), define

\begin{equation}
\mathcal R(\sigma,\eta)
=
\log\left(
4\sin^2\frac{\sigma-\eta}{2}
\right).
\label{eq:periodic_log}
\end{equation}

\noindent Near the diagonal \(\sigma=\eta\), using
\(\sin z=z+O(z^3)\),

\begin{equation}
\begin{aligned}
\mathcal R(\sigma,\eta)
&=
\log\left[
4\left(
\frac{\sigma-\eta}{2}
+
O((\sigma-\eta)^3)
\right)^2
\right]
\\
&=
2\log|\sigma-\eta|
+
O((\sigma-\eta)^2).
\end{aligned}
\label{eq:periodic_log_asymptotic}
\end{equation}

\noindent On the other hand, for the hyperbolic distance in the Poincaré disk
\eqref{eq:hyper_distance}, we write \(r=|x-y|\). Using the conformal factor
\(\lambda(x)=2/(1-|x|^2)\), one finds

\begin{equation}
\log\chi
=
\log r
+
\frac12\log\lambda(x)
+
\frac12\log\lambda(y)
+
O(r^2).
\label{eq:chi_log_expansion}
\end{equation}

\noindent Now let

\[
x=x(t(\sigma)),
\quad
y=x(t(\eta)),
\quad
\eta=\sigma+\tau,
\]

\noindent and consider the limit \(\tau\to0\). Since \(x(t(\sigma))\) is smooth, a
Taylor expansion about \(\sigma\) gives

\begin{equation}
x(t(\sigma+\tau))
=
x(t(\sigma))
+
\frac{d}{d\sigma}x(t(\sigma))\,\tau
+
O(\tau^2),
\end{equation}

\noindent and therefore

\begin{equation}
r
=
|x(t(\sigma))-x(t(\eta))|
=
\left|
\frac{d}{d\sigma}x(t(\sigma))
\right|
\,|\tau|
+
O(\tau^2).
\label{eq:r_local_expansion}
\end{equation}

\noindent Moreover, using \(\eta=\sigma+\tau\) and
\(\sin(\tau/2)=\tau/2+O(\tau^3)\) in \eqref{eq:periodic_log_asymptotic},

\begin{equation}
\frac12\mathcal R(\sigma,\eta)
=
\log|\tau|
+
O(\tau^2).
\label{eq:R_local_expansion}
\end{equation}

\noindent Substituting ~\eqref{eq:R_local_expansion} into
~\eqref{eq:chi_log_expansion}, and absorbing the smooth terms
\(\frac12\log\lambda(x)\), \(\frac12\log\lambda(y)\), together with the
higher-order terms in \(\tau\), into a smooth remainder \(C(\sigma,\eta)\),
gives

\begin{equation}
\log\frac{\chi(\sigma,\eta)}{2}
=
\frac12\mathcal R(\sigma,\eta)
+
C(\sigma,\eta),
\label{eq:log_chi_kress_decomp}
\end{equation}

\noindent where \(C(\sigma,\eta)\) is smooth in a neighbourhood of the diagonal
\(\sigma=\eta\). Using ~\eqref{eq:log_chi_kress_decomp}, the logarithmic part in
~\eqref{eq:kernel_decomposed} becomes

\begin{equation}
\begin{aligned}
-\frac{1}{\pi}
P_\nu'(\cosh\chi)\partial_{n_\eta}\chi
\log\frac{\chi}{2}
={}&
-\frac{1}{2\pi}
P_\nu'(\cosh\chi)\partial_{n_\eta}\chi\,
\mathcal R(\sigma,\eta)
\\
&-
\frac{1}{\pi}
P_\nu'(\cosh\chi)\partial_{n_\eta}\chi\,
C(\sigma,\eta).
\end{aligned}
\label{eq:log_term_kress_substitution}
\end{equation}

\noindent The first term in ~\eqref{eq:log_term_kress_substitution} is the only part
multiplying the explicit Kress logarithm. This is $L_1$ we sought in \eqref{eq:kress_split_goal}

\begin{equation}
L_1(\sigma,\eta)
=
-\frac{1}{2\pi}
P_\nu'\!\bigl(\cosh\chi(\sigma,\eta)\bigr)
\partial_{n_\eta}\chi(\sigma,\eta).
\label{eq:L1_def}
\end{equation}

All remaining terms are collected into the smooth remainder. Equivalently, for
\(\sigma\ne\eta\),

\begin{equation}
L_2(\sigma,\eta)
=
K(\sigma,\eta)
-
L_1(\sigma,\eta)\mathcal R(\sigma,\eta).
\label{eq:L2_offdiag_def}
\end{equation}

\noindent Here \(n_\eta\) denotes the Euclidean outward source normal at
\(y=x(t(\eta))\). Since \(P_\nu'(\cosh\chi)=O(\chi)\) \cite{NIST:DLMF} and
\(\partial_{n_\eta}\chi=O(\chi)\) \eqref{eq:hyper_distance_normal_order}, one has \(L_1(\sigma,\sigma)=0\).

\noindent It remains to compute \(L_2(\sigma,\sigma)\). For small $\chi$ we have

\begin{equation}
P_\nu(\cosh\chi)=1+O(\chi^2).
\label{eq:P_near_zero}
\end{equation}

\noindent Therefore the inverse-\(\chi\) term in ~\eqref{eq:kernel_decomposed} gives after inserting \eqref{eq:P_near_zero}

\begin{equation}
-\frac{1}{\pi}
\frac{P_\nu(\cosh\chi)}{\chi}
\partial_{n_y}\chi
=
-\frac{1}{\pi}
\partial_{n_y}\log\chi
+
O(\chi)\partial_{n_y}\chi .
\label{eq:singular_term_expansion}
\end{equation}

\noindent Since \(\partial_{n_y}\chi=O(\chi)\) \eqref{eq:hyper_distance_normal_order} along a smooth boundary, the second term is
smooth and vanishes on the diagonal. Hence the diagonal value of \(L_2\) is
determined by the finite limit

\begin{equation}
L_2(\sigma,\sigma)
=
-\frac{1}{\pi}
\lim_{\eta\to\sigma}
\partial_{n_\eta}\log\chi(\sigma,\eta).
\label{eq:L2_diag_from_logchi}
\end{equation}

We now compute the boundary limit of \(\partial_{n_y}\log\chi\). Fix a target
point \(x=x(t)\) and let the source point be \(y=x(s)\). From
~\eqref{eq:chi_log_expansion},

\begin{equation}
\begin{aligned}
\partial_{n_x(s)}\log\chi(x(t),x(s))
={}&
\partial_{n_x(s)}\log|x(s)-x(t)| \\
&+
\frac12\partial_{n_x(s)}\log\lambda(x(s))
+
O(|x(s)-x(t)|).
\end{aligned}
\label{eq:normal_log_chi_expansion}
\end{equation}

\noindent The first term in ~\eqref{eq:normal_log_chi_expansion} is the standard
Euclidean diagonal limit. Let \(x(t)\) be a smooth boundary curve and outward unit normal \(n\). Writing \(s=t+\tau\), a Taylor
expansion gives

\begin{equation}
x(s)-x(t)
=
\tau x'(t)
+
\frac{\tau^2}{2}x''(t)
+
O(\tau^3).
\label{eq:boundary_taylor}
\end{equation}

\noindent Using the Frenet decomposition of \(x''(t)\) \cite{frennet}, one finds

\begin{equation}
n(s)\cdot (x(s)-x(t))
=
\frac12 \kappa_E(t)\,|x'(t)|^2\tau^2
+
O(\tau^3),
\label{eq:normal_chord_projection}
\end{equation}

\noindent where $\kappa_E(t)$ is the Euclidean curvature and the $O(\tau)$ term vanishes as it contains the tangent which is orthogonal to the normal. Squaring \eqref{eq:boundary_taylor} gives

\begin{equation}
|x(s)-x(t)|^2
=
|x'(t)|^2\tau^2
+
O(\tau^3).
\label{eq:chord_squared}
\end{equation}

\noindent Since the normal derivative is \(\partial_n f=n\cdot\nabla f\) we have

\begin{equation}
\partial_{n_x(s)}\log|x(s)-x(t)|
=
n(s)\cdot
\frac{x(s)-x(t)}
{|x(s)-x(t)|^2},
\label{eq:normal_derivative_log_chord}
\end{equation}

\noindent division of Eqs.~\eqref{eq:normal_chord_projection} and
\eqref{eq:chord_squared} yields

\begin{equation}
\lim_{s\to t}
\partial_{n_x(s)}\log|x(s)-x(t)|
=
\frac{\kappa_E(t)}{2}.
\label{eq:euclidean_log_limit}
\end{equation}

\noindent Combining Eqs.~\eqref{eq:normal_log_chi_expansion} and
\eqref{eq:euclidean_log_limit}, we obtain

\begin{equation}
\lim_{s\to t}
\partial_{n_x(s)}\log\chi(x(t),x(s))
=
\frac12
\left[
\kappa_E(t)
+
\partial_n\log\lambda(x(t))
\right].
\label{eq:hyperbolic_log_limit}
\end{equation}

\noindent Therefore the diagonal value of the smooth Kress remainder is

\begin{equation}
L_2(\sigma,\sigma)
=
-\frac{1}{2\pi}
\left[
\kappa_E(t(\sigma))
+
\partial_n\log\lambda(x(t(\sigma)))
\right].
\label{eq:L2_diag_compact}
\end{equation}

\noindent Using the conformal factor of the Poincaré disk metric \eqref{eq:poincare_metric},

\begin{equation}
\log\lambda(x)
=
\log 2-\log(1-|x|^2),
\label{eq:lambda_loglambda}
\end{equation}

\noindent one finds

\begin{equation}
\nabla\log\lambda(x)
=
-\frac{\nabla(1-|x|^2)}{1-|x|^2}
=
\frac{2x}{1-|x|^2}.
\label{eq:grad_loglambda}
\end{equation}

\noindent From the definition of the normal derivative (see \eqref{eq:normal_derivative_log_chord}) it follows that

\begin{equation}
\partial_n\log\lambda(x)
=
n\cdot\nabla\log\lambda(x)
=
\frac{2\,x\cdot n}{1-|x|^2}.
\label{eq:dn_log_lambda}
\end{equation}

\noindent The explicit diagonal formula is finally

\begin{equation}
L_2(\sigma,\sigma)
=
-\frac1{2\pi}
\left[
\kappa_E(t(\sigma))
+
\frac{2x(t(\sigma))\cdot n(t(\sigma))}
{1-|x(t(\sigma))|^2}
\right].
\label{eq:L2_diag}
\end{equation}

\noindent Summarizing, for \(\sigma\ne\eta\), the $L_2(\sigma,\eta)$ is evaluated as

\begin{equation}
\begin{aligned}
L_2(\sigma,\eta)
={}&
\frac{1}{\pi}
Q_\nu'\!\bigl(\cosh\chi(\sigma,\eta)\bigr)\,
\partial_{n_\eta}\chi(\sigma,\eta)
\\
&+
\frac{1}{2\pi}
P_\nu'\!\bigl(\cosh\chi(\sigma,\eta)\bigr)\,
\partial_{n_\eta}\chi(\sigma,\eta)\,
\mathcal R(\sigma,\eta),
\end{aligned}
\label{eq:final_kernel_offdiag}
\end{equation}

\noindent with the diagonal $L_2(\sigma,\sigma)$ being \eqref{eq:L2_diag}. Finally from $L_1$ \eqref{eq:L1_def} and $L_2$ \eqref{eq:final_kernel_offdiag} we now evaluate \eqref{eq:kress_split_goal}.

\subsection{Nyström discretization and corner grading}
\label{app:nystrom_kress}

The Kress quadrature is applied in the computational variable \(\sigma\). We use
the shifted equispaced grid

\begin{equation}
\sigma_j=h(j-\frac12),
\quad
h=\frac{2\pi}{N},
\quad
j=1,\ldots,N .
\label{eq:kress_nodes}
\end{equation}

\noindent For a smooth boundary, \(t_j=\sigma_j\). For a piecewise-smooth boundary with
corner grid locations \(c_\alpha\), we use a monotone grading map \(t=t(\sigma)\).
On the interval \([c_\alpha,c_{\alpha+1}]\), define (same idea as in \cite{kress})

\begin{equation}
u(\sigma)=\frac{\sigma-c_\alpha}{c_{\alpha+1}-c_\alpha},
\quad
v_q(u(\sigma))=\frac{u(\sigma)^q}{u(\sigma)^q+(1-u(\sigma))^q},
\label{eq:kress_smoothstep}
\end{equation}

\noindent with $v_q^{'}(0)=v_q^{'}(1)=0$ and set

\begin{equation}
t(\sigma)
=
c_\alpha+(c_{\alpha+1}-c_\alpha)v_q(u(\sigma)),
\quad
q>1.
\label{eq:corner_grading_interval}
\end{equation}

\noindent This map fixes the corners and satisfies \(dt/d\sigma\to0\) at the adjacent
corner endpoints. The physical boundary nodes are

\begin{equation}
x_j=x(t_j),
\quad
t_j=t(\sigma_j),
\label{eq:physical_nodes}
\end{equation}

\noindent and the Euclidean arclength weights are

\begin{equation}
ds_j
=
\left|
\frac{d}{d\sigma}x(t(\sigma_j))
\right|h
=
|x_t(t_j)|\,t'(\sigma_j)h.
\label{eq:graded_arclength_weight}
\end{equation}

\noindent The periodic logarithmic kernel admits the Fourier expansion \cite{kress,kress_book}

\begin{equation}
\mathcal R(\sigma,\eta)
=
\log\left(
4\sin^2\frac{\sigma-\eta}{2}
\right)
=
-2\sum_{m=1}^{\infty}
\frac{\cos\!\bigl(m(\sigma-\eta)\bigr)}{m}.
\label{eq:periodic_log_fourier}
\end{equation}

\noindent For the equispaced computational grid \eqref{eq:kress_nodes} the operator associated with \(\mathcal R\) is represented by the Kress quadrature matrix \(R_{ij}\), which is for even $N$ \cite{kress,fredholm_determinant_method}

\begin{equation}
R_{ij}
=
-\frac{4\pi}{N}
\left[
\sum_{m=1}^{N/2-1}
\frac{\cos\!\left(2\pi m(i-j)/N\right)}{m}
+
\frac{1}{N}(-1)^{i-j}
\right].
\label{eq:kress_matrix_entries_grid}
\end{equation}

\noindent Since \(\sigma_i-\sigma_j=2\pi(i-j)/N\), the matrix depends only on the index
difference \(i-j\) and is therefore circulant. In the implementation, the first
column is constructed from the Fourier coefficients \(1/m\) using an inverse
FFT, and the remaining columns are obtained by cyclic shifts. Thus the Kress
matrix is assembled without ever evaluating the logarithm in
~\eqref{eq:periodic_log_fourier} pointwise. For corner-graded boundaries we use the corresponding graded Kress matrix,
constructed in the same computational variable \(\sigma\) \cite{kress}. The pointwise
logarithm \(\mathcal R(\sigma_i,\sigma_j)\) is used only in the subtraction that
defines \(L_2\); the singular integral itself is evaluated with the matrix
\(R_{ij}\).

\noindent Since \(\sigma_i-\sigma_j=2\pi(i-j)/N\), the smooth-boundary Kress matrix depends only on the index difference \(i-j\) and is therefore circulant. In the implementation, the first column is constructed from the Fourier coefficients \(1/m\) using an inverse FFT, and the remaining columns are obtained by cyclic shifts. Thus the singular quadrature matrix is assembled without evaluating the logarithm pointwise. For corner-graded boundaries we use the corresponding graded Kress matrix in the computational variable \(\sigma\). The grading enters through the mapped boundary points \(x(t(\sigma_j))\), the arclength weights \(ds_j\), and the values of \(L_1\) and \(L_2\). Pointwise values of \(\mathcal R(\sigma_i,\sigma_j)\) are used only off the diagonal in the subtraction that defines \(L_2\); the singular integral itself is evaluated with the matrix \(R_{ij}\).

The logarithmic part is discretized as

\begin{equation}
\int
\mathcal R(\sigma_i,\eta)
L_1(\sigma_i,\eta)\phi(\eta)
\left|\frac{dx}{d\eta}\right|\,d\eta
\approx
\sum_{j=1}^N
R_{ij}L_1(\sigma_i,\sigma_j)
\frac{ds_j}{h}\phi_j.
\label{eq:log_part_discrete}
\end{equation}

\noindent The smooth part is discretized by the trapezoidal quadrature rule,

\begin{equation}
\int
L_2(\sigma_i,\eta)\phi(\eta)
\left|\frac{dx}{d\eta}\right|\,d\eta
\approx
\sum_{j=1}^N
L_2(\sigma_i,\sigma_j)\phi_j ds_j .
\label{eq:smooth_part_discrete}
\end{equation}

\noindent Hence the jump-scaled double-layer matrix used in the computation is

\begin{equation}
D_{ij}
=
R_{ij}L_1(\sigma_i,\sigma_j)\frac{ds_j}{h}
+
ds_jL_2(\sigma_i,\sigma_j),
\label{eq:nystrom_matrix}
\end{equation}

\noindent with ~\eqref{eq:L2_diag} used when \(i=j\). The Fredholm matrix is

\begin{equation}
A(k)=I-D(k).
\label{eq:fredholm_matrix}
\end{equation}

\begin{figure}
    \centering
    \includegraphics[width=\linewidth]{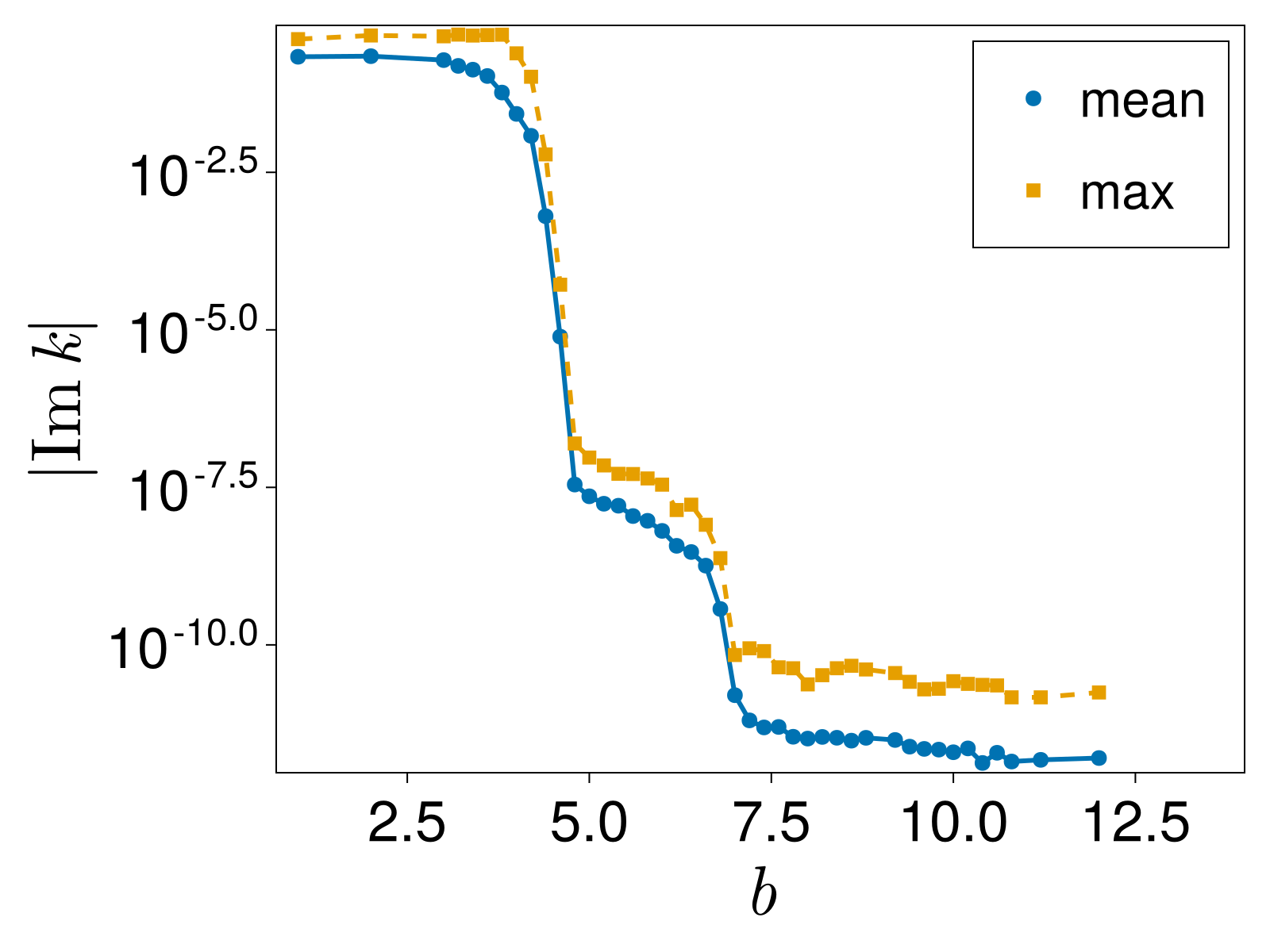}
    \caption{
    Mean and maximum imaginary parts (proxy for eigenvalue accuracy, see Fig.\ref{fig:circle_im_accuracy_convergence}) of the computed eigenvalues in the interval
    $1000\le k\le1010$ for the Schmit triangle, obtained using the Kress quadrature
    with corner grading ($q=2$ in \eqref{eq:kress_smoothstep}), as a function of the points--per--wavelength parameter
    $b$. The vertical axis is logarithmic. The imaginary parts decrease rapidly as
    $b$ is increased, but the convergence is not purely exponential over the full
    range, reflecting the presence of corners and the use of a graded mesh (non--smooth geometry). For
    $b\gtrsim7.5$, the mean imaginary part reaches a plateau near the double-precision
    accuracy floor, while the maximum imaginary part remains slightly larger due to
    the most difficult eigenvalues in the interval. Observations show that with increasing $k$ the needed $b$ for reaching the accuracy saturation plateau slightly decreases.
    }
    \label{fig:schmit_kress_convergence}
\end{figure}

In the computations reported here we used \(q=2\) for the grading parameter.
The number of boundary nodes was chosen from a points-per-wavelength rule based
on the hyperbolic boundary length. We used a discretization density of \(8\) points per
wavelength for the lower part of the spectrum and \(7\) points per wavelength
for the highest eigenvalues (see Fig.\ref{fig:schmit_kress_convergence}). Due to the curvature of the metric, a globally uniform discretization with respect to hyperbolic arclength would require a position-dependent Euclidean point density. However, the Schmit triangle occupies only a small portion of the Poincaré disk, extending to a maximal Euclidean radius of \(r_{\max}\approx0.406\). The corresponding factor $\lambda(r)=\frac{2}{1-r^2}$ therefore varies only from \(\lambda(0)=2\) at the origin to \(\lambda(r_{\max})\approx2.394\), a change of approximately \(20\%\) across the entire billiard. In numerical experiments, enforcing a globally metric-uniform discretization yielded no noticeable improvement in accuracy (see Table.\ref{tab:k3000_kress_gl64} where Gauss--Legendre logarithmic product quadrature used metric scaled points per wavelength ($b$) for each panel length). Moreover, the point furthest from the origin is a corner, where the Kress grading already provides enhanced local resolution.

\begin{table}[htbp]
\centering
\caption{
Comparison of eigenvalues near \(k=3000\) obtained using the Kress split and a
64th-order Gauss--Legendre logarithmic product quadrature (where 256-bit precision was used to get the weights via the left solve of the transposed Vandermonde system - see \cite{Gautschi_Vandermonde} and our implementation \cite{QuantumBilliards_Orel}). The Fredholm matrix was constructed at
\(b=14.0\).
}
\label{tab:k3000_kress_gl64}
\begin{tabular}{ccc}
\hline
\(k_{\mathrm{Kress}}\) &
\(k_{\mathrm{GL64}}\) &
\(|k_{\mathrm{Kress}}-k_{\mathrm{GL64}}|\) \\
\hline
3000.009096186517 & 3000.0090961865135 & \(3.638\times10^{-12}\) \\
3000.0154408337066 & 3000.0154408337200 & \(1.319\times10^{-11}\) \\
3000.0348845105264 & 3000.0348845104477 & \(7.867\times10^{-11}\) \\
3000.0376679889487 & 3000.0376679889127 & \(3.593\times10^{-11}\) \\
3000.0847745340520 & 3000.0847745340790 & \(2.683\times10^{-11}\) \\
3000.1114025142880 & 3000.1114025142733 & \(1.455\times10^{-11}\) \\
3000.1155062578990 & 3000.1155062578660 & \(3.320\times10^{-11}\) \\
3000.1308073911287 & 3000.1308073910870 & \(4.184\times10^{-11}\) \\
3000.1342568665380 & 3000.1342568665104 & \(2.774\times10^{-11}\) \\
3000.1673846791273 & 3000.1673846790650 & \(6.230\times10^{-11}\) \\
3000.1738466768334 & 3000.1738466768047 & \(2.865\times10^{-11}\) \\
3000.1886435077136 & 3000.1886435077380 & \(2.456\times10^{-11}\) \\
\hline
\end{tabular}
\end{table}

\subsection{Evaluation of \(Q_\nu(\cosh\chi)\), \(P_\nu(\cosh\chi)\), and their derivatives}
\label{app:legendre_tables}

The hyperbolic Green function and the Kress logarithmic coefficient require many
evaluations of

\begin{equation}
Q_\nu(\cosh\chi),
\quad
P_\nu(\cosh\chi),
\end{equation}

\noindent together with their derivatives with respect to the hyperbolic distance \(\chi\),
where

\[
\nu=-\frac12+ik.
\]

\noindent Direct special-function evaluation at every matrix entry would be prohibitively
expensive. We therefore evaluate these functions using precomputed Taylor-patch
tables in the radial variable \(\chi\). Let

\begin{equation}
u(\chi)
=
Q_\nu(\cosh\chi),
\quad
y(\chi)
=
\frac{d}{d\chi}Q_\nu(\cosh\chi).
\label{eq:Q_table_variables}
\end{equation}

\noindent For \(u(\chi)=Q_\nu(\cosh\chi)\), the defining Legendre differential equation becomes

\begin{equation}
u'(\chi)=y(\chi),
\quad
y'(\chi)
=
\nu(\nu+1)u(\chi)
-
\coth\chi\,y(\chi).
\label{eq:legendre_radial_system}
\end{equation}

\noindent The same radial system is used for

\[
P_\nu(\cosh\chi) \quad \text{and} \quad \frac{d}{d\chi}P_\nu(\cosh\chi)
\]

\noindent For each fixed \(k\), the interval
\([\chi_{\min},\chi_{\max}]\) is divided into uniform patches centred at
\(\chi_j\). On each patch we write

\begin{equation}
u(\chi_j+\delta)
=
\sum_{m=0}^{p}
a_m^{(j)}\delta^m,
\quad
y(\chi_j+\delta)
=
\sum_{m=0}^{p}
b_m^{(j)}\delta^m.
\label{eq:taylor_patch_form}
\end{equation}

\noindent The coefficients are generated by substituting these expansions into
~\eqref{eq:legendre_radial_system}. If

\begin{equation}
\coth(\chi_j+\delta)
=
\sum_{m=0}^{p}
c_m^{(j)}\delta^m,
\label{eq:coth_patch_series}
\end{equation}

\noindent then the recurrence relations are

\begin{equation}
a_{m+1}^{(j)}
=
\frac{b_m^{(j)}}{m+1},
\quad
b_{m+1}^{(j)}
=
\frac{
\nu(\nu+1)a_m^{(j)}
-
\sum_{\ell=0}^{m}
c_{m-\ell}^{(j)}b_\ell^{(j)}
}
{m+1}.
\label{eq:taylor_patch_recurrence}
\end{equation}

This reduces the expensive part of the computation from one high-precision
special-function evaluation per matrix entry to one high-precision seed per
Beyn quadrature point. Thus, for a Beyn contour with \(N_q\) quadrature points
and an \(N\times N\) boundary matrix, the number of \texttt{mpmath}
high-precision evaluations \cite{mpmath} is reduced from
\(O(N_qN^2)\) to \(O(N_q)\) (see Appendix~\ref{app:beyn_method}). The
remaining work is performed using inexpensive double-precision Taylor
recurrences and Horner evaluation. This avoids repeated high-precision
special-function calls during matrix assembly. This rapid evaluation scheme is also useful e.g. for the evaluation of Tricomi's (confluent hypergeometric) $U$ function appearing in the Green function of a particle in a constant magnetic field (see \cite{magnetic_billiard}).

For very small distances, \(\chi<\chi_0\), the patch tables are replaced by a
local expansion. Using the logarithmic decomposition \eqref{eq:Q_split} we write

\begin{equation}
P_\nu(\cosh\chi)
=
\sum_{m=0}^{M}
\alpha_m\chi^{2m},
\quad
B_\nu(\chi)
=
\sum_{m=0}^{M}
\beta_m\chi^{2m}.
\label{eq:frobenius_series}
\end{equation}

\noindent Only even powers appear because both
\(P_\nu(\cosh\chi)\) and the regular remainder \(B_\nu(\chi)\)
are analytic functions of \(\cosh\chi\) near \(\chi=0\).
Since \(\cosh\chi\) is an even function of \(\chi\), their Taylor expansions
contain only even powers of \(\chi\), while their derivatives contain only odd
powers. The coefficients are generated from the same radial equation using the
series expansion of \(\coth\chi-1/\chi\).

\begin{figure}[h!]
    \centering
    \includegraphics[width=\linewidth]{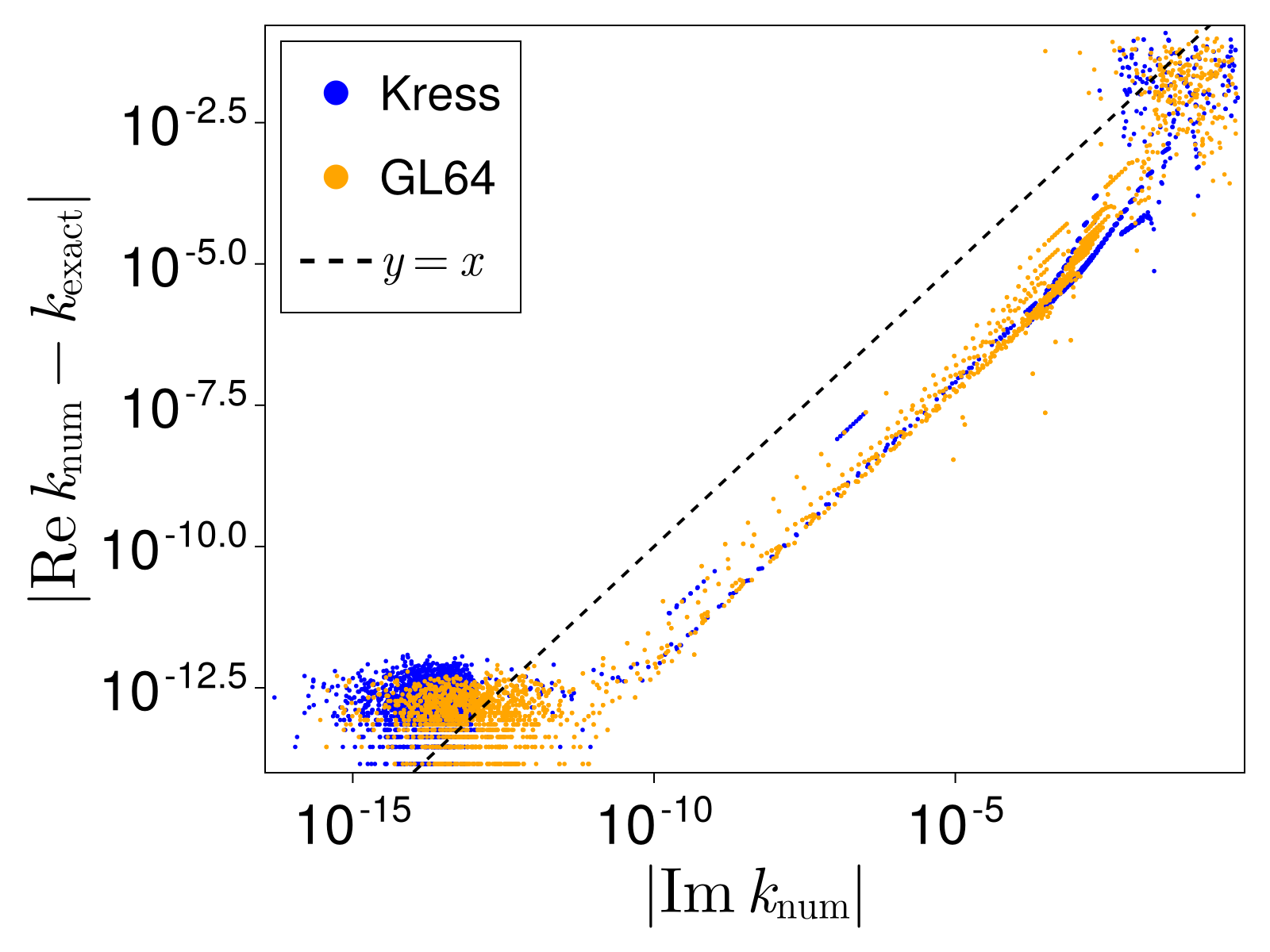}
    \caption{Validation of the hyperbolic boundary-integral discretizations on the circle billiard of radius $R=0.5$. Eigenvalues were computed in the interval $100 \le k \le 120$ for points-per-wavelength parameters $1 \le b \le 10$. The absolute error in the computed eigenvalue real part, $|\operatorname{Re}k_{\mathrm{num}}-k_{\mathrm{exact}}|$, is plotted against the magnitude of the imaginary part, $|\operatorname{Im}k_{\mathrm{num}}|$, for both the Kress and GL64 logarithmic quadratures. The circle reference values $k_{\mathrm{exact}}$ were obtained by solving $P_{-1/2+ik}^{m}(\cosh\rho_R)=0$ in 132 bit precision (where $P$ is the associated Legendre function of the first kind), with $\rho_R=2\operatorname{atanh}R$ (see Appendix.A of \cite{circle_eigvals}). Over many orders of magnitude, the eigenvalue error closely follows the imaginary part, demonstrating that $|\operatorname{Im}k|$ serves as an effective \emph{a posteriori} error indicator. Both methods reach a plateau near $10^{-13}$--$10^{-14}$, beyond which further reduction of the discretization error does not improve the eigenvalue accuracy. This saturation is expected: independent validation of our Legendre-$Q_\nu$ Taylor tables and their derivatives (see Appendix.\ref{app:legendre_tables}) yields a maximum relative error of approximately $2\times10^{-13}$ compared to independent high--precision \texttt{mpmath} evaluations \cite{mpmath}. Consequently, the overall solver accuracy becomes limited by the special-function evaluations rather than by the boundary-integral discretization itself.}
    \label{fig:circle_im_accuracy_convergence}
\end{figure}

In the computations reported here we used a patch spacing of \(10^{-5}\),
Taylor order \(p=8\), and a small-distance threshold
\(\chi_0=10^{-3}\). The series \eqref{eq:frobenius_series} was truncated after
\(24\) even-power coefficients. The same procedure yields
\(P_\nu(\cosh\chi)\) and \(\frac{d}{d\chi}P_\nu(\cosh\chi)\),
which enter the logarithmic coefficient in
~\eqref{eq:L1_def}. The accuracy of the Kress and Gauss--Legendre quadratures was
validated on the exactly separable hyperbolic circle billiard, as shown in
Fig.~\ref{fig:circle_im_accuracy_convergence}.

\subsection{Beyn's contour method}
\label{app:beyn_method}

The boundary integral discretization gives a linear (nonlinear in $k$) problem

\begin{equation}
A(k)u=0,
\label{eq:beyn_nep}
\end{equation}

\noindent where \(A(k)\) is the Fredholm matrix assembled from the hyperbolic
double-layer operator. We solve this problem with Beyn's contour integral
method \cite{beyn_contour}. A detailed description of the implementation was given in ~\cite{c3_arxiv_orel}; here we only recall the steps relevant for the
present computations.

\noindent Let \(\Gamma\) be a positively oriented contour in the complex \(k\)-plane, and
let \(V\in\mathbb C^{N\times r}\) be a random probing matrix. Beyn's method
forms the moment matrices

\begin{equation}
A_0
=
\frac{1}{2\pi i}
\oint_\Gamma
A(z)^{-1}V\,dz,
\quad
A_1
=
\frac{1}{2\pi i}
\oint_\Gamma
zA(z)^{-1}V\,dz.
\label{eq:beyn_moments}
\end{equation}

In practice we use circular or elliptical contours and approximate these
integrals by the trapezoidal rule. For a circular contour
\(z_j=k_0+Re^{i\theta_j}\), with
\(\theta_j=2\pi(j-\frac12)/n_q\), the quadrature weights are $\omega_j=\frac{R}{n_q}e^{i\theta_j}$.

Thus each quadrature point requires assembling and factorizing \(A(z_j)\), after
which the linear systems $A(z_j)X_j=V$ are solved and accumulated into \(A_0\) and \(A_1\). An SVD of $A_0=U\Sigma W^*$ determines the numerical rank \(m\). The integer \(m\) is the number of
eigenvalues inside the contour, counted with multiplicity, provided the probing
dimension \(r\) is sufficiently large. In finite precision, \(m\) is chosen as
the number of singular values of \(A_0\) above a prescribed cutoff. Keeping these
\(m\) singular values gives \(U_m\), \(W_m\), and \(\Sigma_m\). The nonlinear problem is
then reduced to the small matrix

\begin{equation}
B =
U_m^*A_1W_m\Sigma_m^{-1}.
\label{eq:beyn_reduced_matrix}
\end{equation}

\noindent The eigenvalues of \(B\) are the candidate poles inside \(\Gamma\). The corresponding boundary vectors are reconstructed from the reduced eigenvectors \(y_j\) as $u_j=U_m y_j$. Candidates outside the contour are discarded. The remaining candidates are
validated by evaluating the residual $\|A(k_j)u_j\|_2$ and retaining only those below the prescribed tolerance. The real interval \([k_1,k_2]\) is covered by a sequence of contours. Their radii are
chosen using the leading Weyl density $\rho(k) \approx \frac{A_{\Omega}}{2\pi}k$, so that each contour contains a prescribed average number of eigenvalues. For more details see \cite{beyn_contour,QuantumBilliards_Orel}

The computational cost is dominated by dense matrix assembly and factorization at the quadrature nodes. The Legendre Taylor tables described in Sec.~\ref{app:legendre_tables} are therefore crucial: for each quadrature point only a small number of high-precision seed evaluations is required, while all matrix entries are filled using double-precision Taylor evaluation. In the computations reported here, each contour was discretized with \(n_q=40\) trapezoidal quadrature points. The probing dimension was chosen as \(r=m_{\mathrm{Beyn}}+50\), where \(m_{\mathrm{Beyn}}\) denotes the expected number of eigenvalues in the contour, with a typical value \(r\simeq 200-250\). The contours were taken with maximum radius \(R=0.5\), and candidates are retained when the normalized residual satisfied

\begin{equation}
\frac{\|A(k_j)u_j\|_2}{\|u_j\|_2}<10^{-9}.
\end{equation} \label{eq:residual_check}

\noindent Since this is a rather expensive verification step we have relied on a more practical but equivalent heuristic criterion based on the following observation: the only spurious or bad solutions of \eqref{eq:beyn_reduced_matrix} are those with a very large imaginary part. Therefore one need only to order the obtained solutions in decreasing imaginary part and check sequentially the residuals of those with the largest imaginary part via \eqref{eq:residual_check} until one gets $N_{correct}$ solutions in a row. Our choice was $N_{correct}=10$ and it produced no spurious or missing levels. On a positive note such spurious solutions are quite rare and are not present for every Beyn contour. 
\end{document}